\documentclass[aps,pra,showpacs,twocolumn,eqsecnum]{revtex4}
\usepackage{epsfig}
\begin{document}

\title
{Programmable quantum state discriminators with simple programs}
\author{J\'{a}nos A. Bergou$^{1}$}
\author{Vladim\'{\i}r Bu\v{z}ek$^{2}$}
\author{Edgar Feldman$^{3}$}
\author{Ulrike Herzog$^{4}$}
\author{Mark Hillery$^{1}$}
\affiliation{$^{1}$Department of Physics and Astronomy, Hunter College of the City
University of New York, 695 Park Avenue, New York, NY 10021, USA}
\affiliation{$^{2}$ Research Center for Quantum Information, Slovak Academy
of Sciences, 845 11 Bratislava, Slovakia}
\affiliation{$^{3}$Department of Mathematics, Graduate Center of the City University of New York, 365 Fifth Avenue, New York, NY 10016, USA}
\affiliation{$^{4}$Institut f\"ur Physik,  Humboldt-Universit\"at
  zu Berlin, Newtonstrasse 15, D-12489 Berlin, Germany}

\date{\today}
\begin{abstract}
We describe a class of programmable devices that can discriminate between two quantum states. We consider two cases. In the first, both states are {\it unknown}. One copy of each of the unknown states is provided as input, or program, for the two program registers, and the data state, which is guaranteed to be prepared in one of the program states, is fed into the data register of the device.  This device will then tell us, in an optimal way, which of the templates stored in the program registers the data state matches. In the second case, we know one of the states while the other is {\it unknown}.  One copy of the unknown state is fed into the single program register, and the data state which is guaranteed to be prepared in either the program state or the known state, is fed into the data register. The device will then tell us, again optimally, whether the data state matches the template or is the known state. We determine two types of optimal devices. The first performs discrimination with minimum error, the second performs optimum unambiguous discrimination. In all cases we first treat the simpler problem of only one copy of the data state and then generalize the treatment to $n$ copies. In comparison to other works we find that providing $n > 1$ copies of the data state yields higher success probabilities than providing $n > 1$ copies of the program states.
\end{abstract}


\maketitle

\section{Introduction}

Quantum state discrimination \cite{springer} is a basic tool for many tasks in
quantum information and quantum communication. In the prototype problem a quantum processor generates a quantum system as its output which is in one of a set of known states but we do not know which and want to determine the actual state. If the possible states are not orthogonal this cannot be done with 100\% probability of success since the cloning of quantum states is impossible. There are two basic strategies to accomplish state discrimination. In the first, every time a measurement is performed we want to identify the state of the output with one of the possible states. Clearly, errors must be permitted and in the error minimizing strategy the optimum measurement is such that the probability of error is minimum. The case of discriminating with minimum error between two possible states was treated in the pioneering work by Helstrom \cite{helstrom}. More recently, the interest was focused on the unambiguous discrimination. In this strategy we are not permitted to make an erroneous identification of the state.  The cost associated with this condition is that sometimes we fail to identify the state altogether. In the optimum strategy the probability of failure is a minimum. The optimal value of the failure probability for two known and equally likely pure states was obtained by Ivanovic, Dieks and Peres (IDP bound, \cite{ivanovic,dieks,peres}). Later Jaeger and Shimony \cite{jaeger} generalized the IDP bound for arbitrary preparation probabilities of the states, i. e. for arbitrary prior probabilities of the two possible states.  

The actual state-distinguishing device for two {\it known} states depends on
the two states, $|\psi_{1}\rangle$ and $|\psi_{2}\rangle$, i. e. these two
states are ``hard wired'' into the machine.  Another approach is to supply the information about
the states to be distinguished as inputs, in particular as quantum inputs.  That is, one
encodes the information about the states one wants to distinguish into a quantum state, which
is then a kind of quantum program, that is sent into the discriminator at the same time as the
particle whose state is to be identified.  The first such device was proposed by
Du\v{s}ek and Bu\v{z}ek \cite{dusek}.  This device distinguishes the two
states $\cos (\phi /2)|0\rangle \pm \sin (\phi /2)|1\rangle$, and the angle $\phi$ is encoded into
a one-qubit program state in a somewhat complicated way.  The performance of
this device is good; it does not achieve the maximum possible success
probability for all input states, but the average value of its success
probability, averaged over the angle $\phi$, is greater than 90\% of the
optimal value.  In a series of recent works Fiur\'{a}\v{s}ek \emph{et al.}
investigated a closely related programmable device that can perform a von
Neumann projective measurement in any basis, the basis being specified by
the program.  Both deterministic and probabilistic approaches were explored
\cite{fiurasek}, and experimental versions of both the state discriminator
and the projective measurement device were realized \cite{sobusta}.
Sasaki \emph{et al.} developed a related device, which they called a quantum
matching machine \cite{sasaki}.  Its input consists of $K$ copies of two
equatorial qubit states, which are called templates, and $N$ copies of
another equatorial qubit state $|f\rangle$.  The device determines to which
of the two template states $|f\rangle$ is closest.  This device does not
employ the unambiguous discrimination strategy, but optimizes an average
score that is related to the fidelity of the template states and $|f\rangle$.
Programmable quantum devices to accomplish other tasks have been
explored by a number of authors \cite{nielsen}-\cite{zhang}.

Recently two of us proposed an approach to a programmable state discriminating machine
in which the program is related in a simple way to the states $|\psi_{1}\rangle$ and $|\psi_{2}\rangle$ that one is trying to distinguish \cite{bergou}.
A motivation for this problem is that the program state may be
the result of a previous set of operations in a quantum information processing
device, and it would be easier to produce a state in which the information
about $|\psi_{1}\rangle$ and $|\psi_{2}\rangle$ is encoded in a simple way
than one in which it is encoded in a more complicated way.  The program is the most elementary possible, it consists of copies of the states one is trying
to distinguish. The device then performs optimally with the given program states or, in other words, it optimally identifies the data state with one of the two unknown 
program states, or reference states, respectively.
Despite the complete lack of classical information about the reference states,
the identification is still possible due to symmetry properties that are intrinsically quantum mechanical, and are similar to those first employed
by Barnett \emph{et al.} \cite{BCJ} for the purpose of comparing unknown states.  

The original results in \cite{bergou} 
were for unambiguous discrimination and for qubit data and 
program states.  They have recently been extended to qudits by Hayashi \emph{et al.}, both for 
optimum unambiguous discrimination \cite{hayashi2} and for minimum error
discrimination \cite{hayashi1}.  Their investigations are restricted to equal prior probabilities, where the data state equally likely matches each one of the program states, but they also dealt with the case in which an arbitrary number of copies is provided for each of the two program states.

In the present paper we generalize the programmable state discriminator introduced in {\cite{bergou}} in several other directions and develop a comparative study of programmable state discriminators based on the two measurement strategies of minimum-error discrimination and 
optimum unambiguous discrimination.  
For this purpose
in Sec. II we first reformulate the problem of programmable state discriminators as a problem of discrimination between two mixed quantum states.
Sec. III is devoted to the case that both of the pure states to be discriminated are unknown so we need a reference state for each. In Part A we treat
the error minimizing version of the programmable state discriminator, considering both a joint
measurement on all three qubits, and also
a measurement prescription that is restricted to two-qubit measurements only. In Part B we rederive the results of Ref. \cite{bergou} for the unambiguous version of the programmable state discriminator partly for comparison's sake but also using the consistent approach based on the equivalent mixed state discrimination problem. It should be noted, in this context, that the results of Ref. \cite{bergou} were obtained in a somewhat {\emph{ad hoc}} manner and the current approach gives a solid foundation to those results. We also compare the optimal probabilities obtained in Parts A and B for the programmable state discriminators based on the two possible strategies. In Sec. IV we fill another gap and show how to construct devices that can optimally discriminate between one known and one unknown state using both minimum-error and optimum unambiguous strategies. That is, we know what $|\psi_1\rangle$ is, but do not know $|\psi_{2}\rangle$. Then we need a reference state only for the unknown state, which constitutes the program in this case.  We can say that this line of investigation characterizes the quality of the source that produces the states to be discriminated, or the quality of our knowledge about the source, respectively.  If both possible states are known (the original IDP and Helstrom problem) 
there is no need for a program, the states are hard wired into the optimal device. If one of the states is known we need a program for the unknown state while the other is hard wired into the device and if both states are unknown we need a program for both.

We also take look at another aspect of the problem. Namely, besides investigating the effect of the source quality on the optimal performance of this family of state discriminating devices, we also investigate the effect of the resources on the performance of these devices. Suppose that instead of one copy of the state to be discriminated we are given $n$ copies, but we still only
possess one copy each of the unknown state(s) as the program state (or none for two known states).  
In Sec. V we therefore generalize the two unknown qubit scenarios of Sec. III for the case when $n$ copies of the input state, and one copy of the program states, are provided. In Sec. VI we provide a similar generalization of the one unknown qubit cases treated in Sec. IV. 
In each of these cases we determine the optimal measurement strategy 
both for minimum error and unambiguous discrimination of the data state.
The devices that accomplish this are programmable, in the first case the
program consists of two qubits, one in $|\psi_{1}\rangle$ and one in
$|\psi_{2}\rangle$, while in the second case the program consists of a
single qubit in the state $|\psi_{2}\rangle$.  Note that in all cases, the
program is extremely simple.  It is what could be called a ``quantum list'',
a set of qubits, one in each of the states to be discriminated, or one each
in some subset of the states to be discriminated. 
In Sec. VII we conclude with a brief discussion of how these results can be used to characterize the preparation quality (source quality) and to quantify the available resources. 

\section{Discrimination of unknown states and its connection to the discrimination of mixed states}

Let us begin by briefly reviewing the problem that was originally addressed in \cite{bergou}. We  consider a system of three qubits, labeled $A$, $B$, and $C$, and assume that
the qubit $A$ is prepared in the state $|\psi_{1}\rangle$, and the qubit $C$ is prepared in the state $|\psi_2\rangle$. Qubit $B$ is guaranteed to be prepared in either $|\psi_{1}\rangle$ or $|\psi_{2}\rangle$, with a probability $\eta_{1}$ of being in $|\psi_{1}\rangle$ and a probability
$\eta_2=1-\eta_1$ of being in $|\psi_{2}\rangle$.  The states $|\psi_{1}\rangle$ or $|\psi_{2}\rangle$ are different and unknown. Our task is to find whether the state of qubit $B$ is $|\psi_{1}\rangle$ or $|\psi_{2}\rangle$.  One way of looking at this problem is to view the qubits $A$ and $C$ as a program.  They are sent into the program register of some
device, called a programmable state discriminator, and the third,
unknown qubit is sent into the data register of this device.  The device then tells us, with an
optimal probability of success, which one of the two program states the
unknown state of the qubit in the data register corresponds to.  We can consider this  problem 
as a task in measurement optimization.  We want to find
an optimal measurement strategy that, with a maximum probability of success,
tells us which one of the two program states, stored in the program
register, matches the unknown  state, stored in the data register.  In \cite{bergou} only unambiguous discrimination was treated, in which the measurement is allowed to return an inconclusive result but never an erroneous
one.  Here we want to investigate the measurement strategy of minimum-error discrimination, as well.
In general, we want to determine the best possible measurement for identifying the state of the qubit $B$. Our task is then reduced to the following measurement optimization problem.
One has two input states
\begin{eqnarray}
|\Psi_{1}\rangle & = & |\psi_{1}\rangle_{A}|\psi_{1}\rangle_{B}
|\psi_{2}\rangle_{C} \ , \nonumber \\
|\Psi_{2}\rangle & = & |\psi_{1}\rangle_{A}|\psi_{2}\rangle_{B}
|\psi_{2}\rangle_{C} \ ,
\end{eqnarray}
where the subscripts $A$ and $C$ refer to the program registers ($A$ contains
$|\psi_{1}\rangle$ and $C$ contains $|\psi_{2}\rangle$), and the subscript $B$
refers to the data register.  Our goal is to optimally distinguish between
these inputs, with respect to some reasonable criteria, keeping in mind that one has no knowledge of $|\psi_{1}\rangle$ and $|\psi_{2}\rangle$ beyond their \emph{a priori} probabilities. 

Assuming the states $|\psi_1\rangle$ and $|\psi_2\rangle$ to be completely unknown, we have to find the measurement strategy that is optimal on average. Thus, we have to take the average of the input with respect to all possible qubit states. The problem is then equivalent to distinguishing between two mixed states, given by the density operators
\begin{eqnarray}
\label{rh1}
\rho_1 = \left \{|\Psi_ {1}\rangle \langle \Psi_ {1}|\right \}_{\rm av} \ , \\
\label{rh2}
\rho_2 = \left \{|\Psi_{2}\rangle \langle \Psi_{2}|\right \}_{\rm av} \ ,
\end{eqnarray}
that occur with the prior probabilities $\eta_1$ and $\eta_2$, respectively.
Any state of a particular qubit ($A$, $B$ or $C$) can be represented using the Bloch parametrization given by $|\psi_{i}\rangle = \cos (\theta_{i}/2) |0\rangle + e^{i\phi_i} \sin (\theta_{i}/2) |1\rangle$ (i=1,2),
with $|0\rangle$ and $|1\rangle$ denoting an arbitrary set of orthonormal basis states. Here $\theta$ and $\phi$ are the polar and azimuthal angle on the Bloch sphere. After performing the averaging with respect to all possible values of $\theta$ and $\phi$ we arrive at
\begin{eqnarray}
\label{rho1a}
\rho_1 &= & \frac{1}{6} P_{AB}^{sym} \otimes I_{C} \ ,\\
\label{rho2a}
\rho_2 &= & \frac{1}{6} I_{A} \otimes P_{BC}^{sym} \ .
\end{eqnarray}
where $P_{AB}^{sym} = \sum_{i=1}^3 |u_i\rangle_{AC} \,_{AC} \langle u_i|$ and $P_{BC}^{sym}=\sum_{i=1}^3 |u_i\rangle_{BC} \,_{BC}\langle u_i|$ are the projectors onto the symmetric subspaces of the corresponding qubits, $AB$ and $BC$, respectively. Here we used the two-qubit basis states
\begin{eqnarray}
\label{u1}
|u_1\rangle_{AB}=|0\rangle_A |0\rangle_B,\
 |u_2\rangle_{AB}=\frac{|0\rangle_A |1\rangle_B
                          +|1\rangle_A |0\rangle_B}{\sqrt{2}}, \ \ \ \\
\label{u2}
|u_3\rangle_{AB}=|1\rangle_A |1\rangle_B,\
 |\bar{u}_2\rangle_{AB}=\frac{|0\rangle_A |1\rangle_B
                          -|1\rangle_A |0\rangle_B}{\sqrt{2}}, \ \ \
\end{eqnarray}
and the analogous expressions for the qubit combination $BC$.
Due to the symmetry of the state $\rho_1$ with respect to interchanging
qubits $A$ and $B$ the antisymmetric state $|u_4\rangle_{AB}$ does not
enter the expression for the density operator. Eqs. (\ref{rho1a}) and (\ref{rho2a}) reduce our state identification problem to the problem of discriminating between these two mixed states.

When one of the two states that we want to distinguish is known, we arrive at a simpler variant of the discrimination problem. We do not need to provide a template for the known state, and one of the program registers, say $A$, can be eliminated from the problem.
It is convenient to define the single-qubit basis states in such a way that the
known pure state serves as one of the basis states, denoted by $|0\rangle$, so $|\psi_{1}\rangle = |0\rangle$. We then have to distinguish two cases, the qubit $B$ is either in the state
$|0\rangle_B$, occurring with the prior probability $\eta_1$, or it is
in the unknown state of the qubit $C$, occurring with the prior
probability $\eta_2$. These two cases correspond to the density operators
\begin{eqnarray}
\label{sigma1k}
\rho^{\prime}_1  &=&   |0\rangle_B {}_{B}\langle 0| \otimes
 \left \{|\psi \rangle_C {}_{C}\langle \psi| \right \}_{\rm av}
       =  \frac{1}{2} |0\rangle_B {}_{B}\langle 0| \otimes I_{C} \nonumber  \\
       {}&=&\frac{1}{2} \left \{|u_{1}\rangle_{BC} \; _{BC}   \langle u_{1}| + 
|v_{2}\rangle_{BC} \; _{BC}  \langle v_{2}| \right \} ,\\
\label{sigma2k}
\rho^{\prime}_2  &=& \left \{|\psi\rangle_B \,|\psi\rangle_C\
              {}_{B}\langle \psi|\, {}_{C}\langle\psi| \right \}_{\rm av}=
 \frac{1}{3} P_{BC}^{sym} .
\end{eqnarray}
Here
\begin{eqnarray}
\label{v2}
 |v_2\rangle_{AB}=|0\rangle_A |1\rangle_B .
 \end{eqnarray}
In addition, we introduce
\begin{eqnarray}
\label{v2bar}
|\bar{v}_2\rangle_{AB}=|1\rangle_A |0\rangle_B ,\ \ \ \
\end{eqnarray}
together with  the analogous expression for the qubit combination $BC$. $\{|u_{2}\rangle,|\bar{u}_{2}\rangle\}$ and$\{|v_{2}\rangle,|\bar{v}_{2}\rangle\}$ form alternative bases for the subspace with exactly one qubit in the state $|1\rangle$. They will prove useful later when we consider the various discrimination scenarios in the following sections.

After these preliminary considerations we are now in a position
to investigate different possible measurements
for identifying the state of the qubit $B$,
 i. e. for distinguishing between the two density operators
given by Eqs. (\ref{rho1a}) and (\ref{rho2a}) or, alternatively, by
Eqs. (\ref{sigma1k}) and (\ref{sigma2k}).
Before doing so, we briefly recall the underlying
theoretical concepts for treating  the strategies
of discriminating two mixed states.
Any measurement suitable for distinguishing between the mixed states
$\rho_1$ and $\rho_2$, occurring with the prior
probabilities $\eta_1$ and $\eta_2=1-\eta_1$, respectively, can be formally
described with the help of three positive detection operators
${\Pi}_0$, ${\Pi}_1$ and  ${\Pi}_2$, whose sum is the identity,
\begin{equation}
 {\Pi_0}+ {\Pi_1} + {\Pi_2} = I.
\label{Pi}
\end{equation}
These operators are defined in such a
way that for $j=1,2$
${\rm Tr}({\rho}{\Pi}_j)$ is the probability to infer from the measurement that
the system
is in the state $\rho_j$ if it has been prepared in a state
$\rho$, while ${\rm Tr}({\rho}{\Pi}_0)$ is the probability
that the measurement result is inconclusive, i. e. that the
measurement fails to give a definite answer.
When all detection
operators are projectors, the measurement is a von Neumann measurement,
otherwise it is
a generalized measurement based on a positive operator-valued measure (POVM).
Once the detection operators of a generalized measurement have been found,
Neumark's theorem guarantees that schemes for actually realizing the
measurement can be devised by determining suitable projections in an enlarged Hilbert
space that results from appending an  ancilla to the original system
\cite{neumark,preskillnotes}.

The above POVM is appropriate for unambiguous state discrimination.
For {\it minimum-error discrimination} inconclusive results do not occur, so that
\begin{equation}
\label{Pi0}
\Pi_0=0,
\end{equation}
and we require that the probability of errors in the discrimination procedure is a minimum.
For two mixed states this problem was originally solved by Helstrom \cite{helstrom}.
The error probability is always larger than zero unless the states to
be distinguished are orthogonal, and it can be expressed as
\begin{eqnarray}
P_{err}&=&
\eta_1{\rm Tr}({\rho}_1{\Pi}_2) +
\eta_2{\rm Tr}({\rho}_2{\Pi}_1)\nonumber\\
 &&= \eta_1 + {\rm Tr}[(\eta_2 \rho_2 - \eta_1 \rho_1)\Pi_1],
 \label{Perr}
\end{eqnarray}
where in the second line Eqs. (\ref{Pi}) and (\ref{Pi0}) have been used,
as well as the relation $\eta_2= 1 - \eta_1$.
After introducing the operator
\begin{equation}
\Lambda= \eta_2 {\rho}_2 - \eta_1 {\rho}_1
= \sum_{k} \lambda_k |\phi_k\rangle \langle \phi_k|
\label{-lambda}
\end{equation}
it is obvious that the minimum of the error probability is obtained
when $\Pi_1$ is the projector onto those eigenstates $|\phi_k\rangle$
of $\Lambda$ that belong to negative eigenvalues $\lambda_k$.
The optimum detection operators therefore read
\begin{equation}
{\Pi}_{1}^{\rm opt}   =  \sum_{k < k_0}
|\phi_k\rangle \langle \phi_k|,
\qquad
{\Pi}_{2}^{\rm opt}   =  \sum_{k \geq k_0}
|\phi_k\rangle \langle \phi_k|,
\label{opt}
\end{equation}
where $\lambda_k   <  0 $ for $1 \leq k < k_0$ and
$\lambda_k   \geq  0 $ for $k \geq k_0$.
Clearly, these two operators are projections, and the optimal minimum-error
measurement for discriminating between two quantum
states is, therefore, always a von Neumann measurement.
The resulting minimum error probability
$P_{err}^{min}=P_E$ is given in \cite{helstrom} by
\begin{equation}
P_E = \frac{1}{2}\left(1 - {\rm Tr} |\eta_2 \rho_2 - \eta_1
{\rho}_1|\right)
=\frac{1}{2} \left( 1 - \sum_{k}|\lambda_k|\right).
\label{hel}
\end{equation}

In {\it optimum unambiguous discrimination} which is the other frequently used strategy errors are not allowed to occur. This requirement is equivalent to
\begin{equation}
\label{Pi1}
\rho_1 \Pi_2 = \rho_2 \Pi_1= 0,
\end{equation}
(see, for example \cite{springer}).
In the optimum measurement scheme the failure probability,
i. e. the probability for getting
an inconclusive outcome, is minimized,
taking into account the constraint that
the eigenvalues of the operator $\Pi_0 = I - \Pi_1 - \Pi_2$
are non-negative.
The failure probability is always nonzero unless the states to be discriminated are
orthogonal. It can be expressed as
\begin{eqnarray}
Q_{fail}&=& \eta_1 \rm Tr (\rho_1\Pi_0) + \eta_2 \rm Tr (\rho_2\Pi_0)\nonumber\\
&&= 1- \eta_1{\rm Tr}({\rho}_1{\Pi}_1) -
\eta_2{\rm Tr}({\rho}_2{\Pi}_2)\nonumber\\
&&=1-P_{succ},
\label{Qfail}
\end{eqnarray}
where  Eqs. (\ref{Pi}) and (\ref{Pi1}) have been used, and where we also introduced the 
success probability $P_{succ}$ of the measurement.
Optimum unambiguous discrimination between two mixed states is an issue
of ongoing theoretical research \cite{raynal,eldar,feng,rudolph,
HB3,raynal1,BFH}.
 In contrast to minimum-error discrimination,
there does not exist a compact formula expressing the minimum probability of inconclusive
results, i. e. the minimum failure probability, for unambiguously discriminating
two mixed states that are completely arbitrary. However,
analytical solutions can be obtained for certain special
classes of density operators, including the cases that are of interest for this paper.

\section{Two-qubit program, single copy of the data state}

\subsection{Minimum-error discrimination strategy}

\subsubsection{Joint measurement on all three qubits}

We begin by investigating the measurement that discriminates,
with minimum probability of error, between the density operators
given by Eqs. (\ref{rho1a}) and (\ref{rho2a}).
For this purpose we define the orthonormal basis states in the eight-dimensional Hilbert space spanned by the three qubits as
\begin{eqnarray}
\label{ba}
|1\rangle &=& |0\rangle_A |0\rangle_B |0\rangle_C, \qquad
|2\rangle = |0\rangle_A |0\rangle_B |1\rangle_C,\nonumber\\
|3\rangle &=& |0\rangle_A |1\rangle_B |0\rangle_C, \qquad
|4\rangle = |1\rangle_A |0\rangle_B |0\rangle_C,\nonumber\\
|5\rangle &=& |0\rangle_A |1\rangle_B |1\rangle_C, \qquad
|6\rangle = |1\rangle_A |0\rangle_B |1\rangle_C,\nonumber\\
|7\rangle &=& |1\rangle_A |1\rangle_B |0\rangle_C, \qquad
\label{be}
|8\rangle = |1\rangle_A |1\rangle_B |1\rangle_C.
\end{eqnarray}
The numbering of the states is essentially the binary number formed by the bit values on the right-hand side shifted by one. Note, however, that in the case of $|4\rangle$ and $|5\rangle$ the order is reversed. Expanding the expressions for $\rho_1$ and $\rho_2$ in this basis and
introducing the notations,
\begin{eqnarray}
|r_1\rangle &=& \frac{|3\rangle + |4\rangle}{\sqrt{2}} , \qquad
|r_2\rangle = \frac{|5\rangle + |6\rangle}{\sqrt{2}} , \\
|s_1\rangle &=& \frac{|2\rangle + |3\rangle}{\sqrt{2}} , \qquad
|s_2\rangle = \frac{|6\rangle + |7\rangle}{\sqrt{2}} ,
\end{eqnarray}
we obtain the spectral representations
\begin{eqnarray}
\label{rho1}
\rho_1 = \frac{1}{6} (\sum_{l=1}^{2}  |r_l\rangle\langle r_l|
+ |1\rangle\langle 1| + |2\rangle\langle 2|
+ |7\rangle\langle 7| + |8\rangle\langle 8|),
\;\;\;\\
\label{rho2}
\rho_2 = \frac{1}{6}(\sum_{l=1}^{2}  |s_l\rangle\langle s_l|
+ |1\rangle\langle 1| + |4\rangle\langle 4| +
|5\rangle\langle 5| + |8\rangle\langle 8|).
\;\;\;
\end{eqnarray}
When we express the operator $\Lambda= \eta_2 {\rho}_2 - \eta_1 {\rho}_1$
with the help of the basis states given by  Eq. (\ref{ba}), we arrive at
an eight-dimensional square matrix which is block-diagonal
if the columns and rows are numbered according to the numbering
of the basis states. It can be written as
\begin{eqnarray}
\label{Lambda}
\Lambda ={ \left( \begin{array}{cccc}
    L_1 &  0   & 0  & 0 \\
      0 &  L_3 & 0  & 0 \\
      0 &  0   & L_3& 0 \\
      0 &  0   & 0  & L_1
         \end{array} \right) },\;\;
\end{eqnarray}
where $L_1 = (\eta_2-\eta_1)/6$ and
\begin{equation}
L_3 = \frac{1}{12}{ \left( \begin{array}{ccc}
         \eta_2-2\eta_1 &   \eta_2      &  0 \\
          \eta_2        & \eta_2-\eta_1 &  -\eta_{1} \\
            0           &   -\eta_{1}   & 2\eta_2-\eta_1
         \end{array} \right) }.
 \end{equation}
Since the eigenvalues of $L_3$ are $(\eta_2-\eta_1)/6$ and
\begin{equation}
\lambda_{\pm}=
\frac{1}{12}\left[\,\eta_2-\eta_1
\pm\sqrt{(\eta_2-\eta_1)^2+3\eta_1\eta_2}\,\right],
\end{equation}
the complete set of eigenvalues of the operator $\Lambda$ is given by
\begin{eqnarray}
\lambda_{1}= \lambda_{2}= \lambda_{-}, \quad \lambda_{3}=\lambda_{4}=\lambda_{+}, \nonumber \\
\lambda_k = \frac{\eta_2-\eta_1}{6} \quad (5\leq k \leq 8).
\end{eqnarray}
The corresponding eigenstates, $|\phi_{k}\rangle$, are found to be
$|\phi_7\rangle = |1\rangle$, $|\phi_8\rangle = |8\rangle$,
\begin{eqnarray}
\label{eig}
|\phi_1\rangle &=& \frac {a^{-}|2\rangle + |3\rangle + b^{-} |4\rangle}
{\sqrt{1 + (a^{-})^2 + (b^{-})^2}} \ , \nonumber \\
|\phi_3\rangle &=& \frac {a^{+}|2\rangle + |3\rangle + b^{+} |4\rangle}
{\sqrt{1 + (a^{+})^2 + (b^{+})^2}} \ , \nonumber \\
|\phi_5\rangle &=& \frac {|2\rangle + |3\rangle + |4\rangle}{\sqrt{3}} \ ,
\end{eqnarray}
where
\begin{eqnarray}
a^{\pm} &=& \pm \sqrt{(\eta_1-\eta_2)^2+3\eta_1\eta_2} - \eta_{1} \ , \nonumber \\
b^{\pm} &=& \mp \sqrt{(\eta_1-\eta_2)^2+3\eta_1\eta_2} - \eta_{2} \ .
\end{eqnarray}
The eigenstates $|\phi_2\rangle$, $|\phi_4\rangle$, $|\phi_6\rangle$ follow
from replacing the ordered set $\{|2\rangle,|3\rangle,|4\rangle\}$ by
the ordered set $\{|5\rangle,|6\rangle,|7\rangle\}$ in the expressions given by
Eq. (\ref{eig}).
By inserting the eigenvalues of $\Lambda$ into  Eq. (\ref{hel})
we find after a little algebra that the minimum probability of error can be written in the following compact way,
\begin{equation}
\label{PE}
P_E= \eta_{min}\left(1 - \frac{1}{2}\frac{\eta_{max}}{\eta_{max}-\eta_{min} + \sqrt{1-\eta_{max}\eta_{min}}}\right) \ ,
\end{equation}
where $\eta_{max}$ ($\eta_{min}$) is the larger (smaller) of $\eta_{1}$ and $\eta_{2}$.
This result is in agreement with the one derived in \cite{hayashi1} for the special case that 
$\eta_{max}=\eta_{min}=1/2$.

The above expression lends itself to a transparent interpretation. The error probability $P_{e}$ would be $\eta_{min}$ if we did not perform any measurement at all but would simply guess, always choosing the state whose \emph{a priori} probability is larger. The factor in the bracket, multiplying $\eta_{min}$, is the improvement due to the optimized measurement. It is a slowly varying function of the prior probabilities, its value lying between $0.71$ and $0.75$ in the entire $0<\eta_{min}<1/2$ interval.
To be specific, let us assume that the qubits are labeled in such a way that
$\eta_1$ is the smaller of the two prior probabilities, i. e. that
$\eta_1\leq 0.5$. In this case
$\lambda_1$ and $\lambda_2$ are the only negative eigenvalues, and
the optimum detection operators ${\Pi}_{1}^{\rm opt}$
and ${\Pi}_{2}^{\rm opt}$ for minimum-error identification take the form
given by Eq. (\ref{opt}) with $k_0=3$.
This means that
the qubit $B$ is inferred to be in the state of qubit $A$ when a projection onto the
subspace spanned by  the eigenstates of ${\Pi}_{1}^{\rm opt}$ is successful,
and after successful projection
onto the complementary subspace it is inferred to be in the state of qubit $C$.

From the structure of the eigenstates $|\phi_k\rangle$ determining the optimum detection
operators, and from the definition of the basis states, given by Eq. (\ref{ba}), it is obvious
that the smallest possible error probability, $P_E$, can only be obtained by performing a joint measurement on all three qubits simultaneously. The question therefore naturally arises as to what is the smallest value of the error probability achievable under the restriction that only joint measurements on two qubits are allowed. In the following we study this problem.   This situation
is worth examining for two reasons.  First, two-qubit measurements are easier to perform than
three-qubit ones.  Second, by comparing the results of the two- and three-qubit measurements,
we see how the additional quantum information contained in the third qubit affects the result.

\subsubsection{Restriction to two-qubit-measurements}

First let us assume that the qubit $C$ is not accessible, but that
we are able to perform a joint measurement on the qubits $A$ and $B$. This would be the
case, for example, if a copy of only one of the two states we are trying to distinguish is
provided.  Starting again from Eqs. (\ref{rho1a}) and (\ref{rho2a}),
the problem  of identifying the state of $B$ is then equivalent to discriminating
between the two reduced density operators,
\begin{eqnarray}
\label{sigma1a}
 {\rm Tr}_C \rho_1 &=& \frac{1}{3} P_{AB}^{sym} \ , \\
\label{sigma2a}
 {\rm Tr}_C \rho_2 &=&
 \frac{1}{4} I_{A} \otimes I_{B} \ ,
\end{eqnarray}
with the prior probabilities $\eta_1$ and $\eta_2$, respectively.

The operator  $\tilde{\Lambda}=\eta_2{\rm Tr}_C \rho_2 - \eta_1{\rm Tr}_C \rho_1$,
relevant for minimum-error discrimination, is now given by
\begin{equation}
 \label{Lambda1}
\tilde{\Lambda} = \left( \frac {\eta_2}{4}- \frac {\eta_1}{3} \right)
   \sum_{i=1}^3 |u_i\rangle_{AB} {}_{AB}\langle u_i| +
    \frac {\eta_2}{4}|\bar{u}_2\rangle_{AB} {}_{AB}\langle \bar{u}_2| \ .
\end{equation}
When $\eta_{1} \leq 3\eta_{2}/4$ (or $\eta_1 \leq 3/7$) all four eigenvalues of $\tilde{\Lambda}$ are positive,
and from Eq. (\ref{opt}) the optimum detection operators are
obtained as $\Pi_2^{\rm opt} =I$ and $\Pi_1^{\rm opt}=0$.
Hence the minimum error probability
is achieved by guessing that the quantum system is always in the
state that is more probable, in this case $|\psi_2\rangle$,
without performing any measurement at all.
This is a special situation, described earlier \cite{hunter},
which has been observed in connection with a different problem
of two-qubit-discrimination \cite{H}.

On the other hand, for $\eta_{1} \geq 3\eta_{2}/4$ (or $\eta_1 \geq 3/7$) we readily find that
$\Pi_2^{\rm opt} = |\bar{u}_2\rangle_{AB} {}_{AB}\langle \bar{u}_2|$
and $\Pi_1^{\rm opt} = \sum_{i=1}^3 |u_i\rangle_{AB} {}_{AB}\langle u_i|$,
i. e. that the error probability is smallest when
the qubit $B$ is guessed to be in the state of qubit $A$ after a successful
projection onto the symmetric subspace of qubits $A$ and $B$, while there is no guessing involved after a successful projection onto the antisymmetric subspace of qubits $A$ and $B$. It is then known with certainty to be not in the state of qubit $A$.

The results for the minimum error probability, following from
Eqs. (\ref{Lambda1}) and (\ref{hel}), can be summarized as
\begin{equation}
\label{PEAB}
P_E^{AB}= \left \{
\begin{array}{ll}
\eta_1  \;\; & \mbox{if $\eta_1 \leq \frac{3}{7}$}\\
\frac{3}{4}(1-\eta_1)
         \;\; & \mbox{otherwise.}
\end{array}
\right.
\nonumber
\end{equation}
Similarly, a joint measurement on the qubits $B$ and $C$
yields the minimum error probability
\begin{equation}
\label{PEBC}
P_E^{BC}= \left \{
\begin{array}{ll}
\frac{3}{4}\eta_1  \;\; & \mbox{if $\eta_1 \leq \frac{4}{7}$}\\
1-\eta_1  \;\; & \mbox{otherwise.}
\end{array}
\right.
\nonumber
\end{equation}
Fig. 1 also reveals that by performing the optimal two-qubit measurement an error probability
can be achieved that is almost as low as the absolute minimum error probability, $P_E$, given by Eq. (\ref{PE}) where the latter can only be reached by a joint measurement on all three qubits.
Even when the advantage of the three-qubit measurement is largest, which happens
for equal prior probabilities, $\eta_{1}=\eta_{2}=1/2$, the difference in the respective minimum error probabilities for state identification is only marginal,
\begin{equation}
\label{PEC}
P_E^{AB}= P_E^{BC}= 0.375, \quad
P_E=\frac{1}{2}-\frac{1}{4\sqrt{3}}=0.356.
\end{equation}

In the next paragraph we compare the minimum probabilities of error with the minimum
probability of failure arising in the other important measurement strategy, that of
unambiguous discrimination.
\begin{figure}[ht]
\begin{center}
\epsfig{file=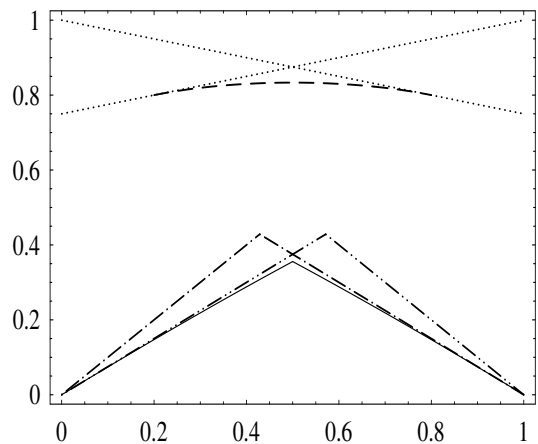,height=6cm,width=7cm}
\caption
{Comparison of the optimum performances of the error minimizing strategy and the unambiguous discrimination strategy for the discrimination of two unknown pure states. The minimum error probabilities $P_E$ resulting from
a three-qubit measurement (full line) and $P_E^{AB}$, $P_E^{BC}$ resulting from two-qubit
measurements (dash-dotted and dash-double dotted line, respectively)
are compared to the failure probabilities for unambiguous discrimination
$Q_F^{AB}$, $Q_F^{BC}$ (dotted lines) and $Q_{F}^{POVM}$ (dashed line).
The error and failure probabilities are plotted vs. the prior probability $\eta_1$ that the state of the data qubit $B$ matches the state of the program qubit $A$.}
\end{center}
\end{figure}

\subsection{Optimum unambiguous discrimination strategy}

The optimum measurement for unambiguously identifying the state
of the data qubit $B$ was found in \cite{bergou} using a method that relied on a
special Ansatz for the detection operators, justified by the symmetry properties of the inputs.
For completeness, here we reconsider the problem in the framework of
the optimum unambiguous discrimination of two mixed states.

In the following we apply the method developed in \cite{HB3} which, in turn, is a special case of the more general approach in \cite{BFH}.
Starting from Eqs. (\ref{rho1}) and (\ref{rho2}), we denote the projectors
onto the  supports of $\rho_1$ and $\rho_2$ by
$P_1$ and $P_2$, respectively. The eigenstates of the operators
$I-P_2$ and $I-P_1$ are easily found to be
\begin{eqnarray}
\label{v}
|a_1\rangle &=& \frac{|2\rangle - |3\rangle} {\sqrt{2}} ,\qquad
|a_2\rangle = \frac{|6\rangle - |7\rangle} {\sqrt{2}},  \\
\label{w}
|b_1\rangle &=& \frac{|3\rangle - |4\rangle} {\sqrt{2}} ,\qquad
|b_2\rangle = \frac{|5\rangle - |6\rangle} {\sqrt{2}} .
\end{eqnarray}
Clearly, $\rho_2 |a_i\rangle=0$ and $\rho_1 |b_i\rangle=0$ for $i=1,2$.
The most general Ansatz for the detection operators,
satisfying $\Pi_1\rho_2= \Pi_2\rho_1=0$ as required
for unambiguous discrimination, therefore reads \cite{HB3}
\begin{equation}
\label{op}
\Pi_1=\sum_{i,j=1}^2\alpha_{ij}|a_i\rangle \langle a_j|,\qquad
\Pi_2=\sum_{i,j=1}^2\beta_{ij}|b_i\rangle \langle b_j|.
\end{equation}
From Eq. (\ref{Qfail})
we readily find that these two detection operators give rise
to the failure probability
\begin{equation}
\label{Qf}
Q_{fail}=
1- \frac{1}{8} \sum_{i=1}^2(\eta_1 \alpha_{ii} + \eta_2 \beta_{ii}).
\end{equation}
Note that due to the  structure of the two given density operators
the failure probability does not depend on the off-diagonal elements of the
detection operators given by Eq. (\ref{op}), a property that is
common to all problems of optimum unambiguous
discrimination of two mixed states that have been explicitly solved so far
\cite{rudolph,HB3,raynal1,BFH}.
We are therefore free to choose $\alpha_{ij}=\beta_{ij}=0$ for $i\neq j$,
a choice that guarantees that $\Pi_0$ is positive for the
largest possible values of $\alpha_{ii}$ and $\beta_{ii}$ $(i=1,2)$, i. e.
that $Q_{fail}$ can be made as small as possible.

When we represent the operator $\Pi_0$ in the basis defined in Eqs. (\ref{ba}), we again arrive at
a block-diagonal eight by eight matrix, similar to Eq. (\ref{Lambda}), given by
\begin{eqnarray}
\Pi_0 ={ \left( \begin{array}{cccc}
        1 & 0 & 0 & 0 \\
        0 &  M(\alpha_{11},\beta_{11}) & 0  & 0 \\
        0 &  0   & M(\beta_{22},\alpha_{22}) & 0 \\
        0 &  0   & 0  & 1
         \end{array} \right) }.\;\;
\end{eqnarray}
Here we introduced the abbreviation
\begin{equation}
M(a,b) = \frac{1}{2}{ \left( \begin{array}{ccc}
          2-a  &   a       &  0  \\
           a   &  2-a-b    &  b  \\
           0   &   b       & 2- b
         \end{array} \right) }.
 \end{equation}
The eigenvalues of $M(a,b)$ are found to be 1 and
\begin{equation}
\mu_{\pm}(a,b) =1-\frac{1}{2}\left(a +
b \pm \sqrt{(a - b)^2+ a b}\right),
\label{mu}
\end{equation}
where obviously $\mu_{\pm}(a,b)=\mu_{\pm}(b,a)$.
All eigenvalues of $\Pi_0$ are nonnegative  provided that
$\mu_{+}(\alpha_{ii},\beta_{ii})$
is nonnegative for $i=1,2$ which holds true when
 $\beta_{ii} \leq (4-4 \alpha_{ii})/(4-3 \alpha_{ii})$.
Hence in order to make $Q_{fail}$ as small as possible,
while keeping $\Pi_0$ a positive operator, we chose the equality sign
and put
\begin{equation}
\beta_{ii} =  \frac {4-4\alpha_{ii}}{4-3\alpha_{ii}}\qquad(i=1,2).
\end{equation}

After inserting these expressions into  Eq. (\ref{Qf}) the resulting
function of $\alpha_{11}$ and $\alpha_{22}$ has to be minimized,
 taking into account that
$0\leq \alpha_{ii} \leq 1$, which
follows from the fact that ${\rm Tr}(\rho\Pi_1)$ describes a probability
for any density operator $\rho$.
In accordance with the optimization problem solved in \cite{bergou},
we find that the failure probability takes its smallest possible value when
$\alpha_{11}=\alpha_{22}=\alpha$, where
\begin{equation}
\label{alpha2}
\alpha= \left \{
\begin{array}{ll}
0  \;\; &  \mbox {if $ \;\;\eta_1 \leq \frac{1}{4}\eta_2 $}  \\
\frac{2}{3}\left(2-\sqrt{\frac{\eta_2}{\eta_1}}\right)
         \;\; & \mbox{if $\;\;\frac{1}{4}\eta_2\leq \eta_1 \leq 4\eta_2$}\\
 1  \;\; & \mbox{if $\;\;\eta_1 \geq 4\eta_2$}.\\
\end{array}
\right.
\nonumber
\end{equation}
Using Eqs. (\ref{op}) and (\ref{ba}) we arrive at the optimum detection operators
\begin{eqnarray}
\label{op6}
\Pi_1^{\rm opt} & = & \alpha\; I_{A} \otimes  |\bar{u}_2\rangle_{BC}\, \langle \bar{u}_2|_{BC}, \\
\Pi_2^{\rm opt}  & = & \frac {4-4\alpha}{4-3\alpha}\;
         |\bar{u}_2\rangle_{AB}\, \langle \bar{u}_2|_{AB} \otimes I_{C} ,
\end{eqnarray}
where the value of $\alpha$ in the different parameter regions for
$\eta_1$ and $\eta_2$ is given by Eq. (\ref{alpha2}). Clearly, in the first
parameter region $\Pi_1^{\rm opt}=0$, while $\Pi_2^{\rm opt}$
describes a projection onto the antisymmetric two-qubit state $|\bar{u}_2\rangle_{AB}$.
Similarly, in the third parameter region a
projection onto the antisymmetric state $|\bar{u}_2\rangle_{BC}$ has to be performed
for optimum unambiguous discrimination.
The failure probabilities resulting from these two von Neumann measurements are
\cite{bergou}
\begin{equation}
\label{neumann}
Q_F^{AB} = 1- \frac{\eta_2}{4} = \frac{3}{4} +\frac{\eta_{1}}{4},\qquad
Q_F^{BC} = 1- \frac{\eta_1}{4}.
\end{equation}
In the intermediate region of the prior probabilities the optimum measurement is
a generalized measurement, yielding the failure probability \cite{bergou}
\begin{equation}
 Q_{F}^{POVM} = \frac{2 + \sqrt{\eta_1(1-\eta_1)}}{3} \quad
 \mbox{$ \left( \frac{1}{5}  \leq \eta_1 \leq \frac{4}{5}\right) $},
\label{POVM}
\end{equation}
where we took into account that $\eta_2=1-\eta_1$.

As seen in Fig. 1, the minimum failure probability is always at least twice as large as the minimum error probability $P_E$ for identifying the qubit state. This agrees with the general relation between the failure probability of optimal unambiguous discrimination and the error probability of minimum-error discrimination of two mixed states that was derived in \cite{HB2}.
Fig. 1 also shows that the advantage of performing a generalized measurement,
as compared to the best projective two-qubit measurement, is small.
For $\eta_1=\eta_2=0.5$, where $Q_F^{AC} = Q_F^{BC} = 7/8 = 0,875$ the failure probability
is only reduced to the value $Q_{F}^{POVM} = 5/6=0.833$.  Of course, the surprise is not the high value of the failure probability but that the success probability is finite for the discrimination of completely unknown states. As we shall see in the next section,
these relatively large failure probabilities are somewhat reduced
when one of the reference states is known.

\section{One-qubit program, single copy of the data state}

\subsection{Minimum-error discrimination between one known and one unknown state}

Now we treat the simplified case in which we want to
decide whether the qubit $B$ is in the known state $|0\rangle_{B}$,
or whether it is in the unknown state of the program qubit stored in register $C$.
We then have to distinguish between the density operators
$\rho^{\prime}_1$ and $\rho^{\prime}_2$ given by
Eqs. (\ref{sigma1k}) and (\ref{sigma2k}) that refer to the first and second alternative, respectively, and that occur with the prior probabilities $\eta_1$ and $\eta_2$.
The subsequent treatment proceeds along exactly the same lines that we followed in  the previous sections.

For minimum-error identification we have to determine the eigenvalues and eigenstates of the

operator  $\Lambda^{\prime} = \eta_2 \rho^{\prime}_2 - \eta_1 \rho^{\prime}_1$.
It is easy to obtain the  spectral representation
\begin{equation}
 \label{Lambda2}
\Lambda^{\prime} = \left( \frac{\eta_2}{3}- \frac {\eta_1}{2} \right)
    |u_1\rangle \langle u_1|  +\frac{\eta_2}{3} |u_3\rangle \langle u_3|
  + \sum_{i=\pm} \lambda_i |\varphi_i\rangle \langle \varphi_i|,
\end{equation}
where
$\lambda_{\pm} = \frac{1}{12}\left(2\eta_2 - 3\eta_1
            \pm \sqrt{4\eta_2^2 + 9 \eta_1^2}\right)$
and
\begin{equation}
 \label{phi}
|\varphi_{\pm}\rangle=\frac{|u_3\rangle-c_{\pm}|\bar{u}_2\rangle}{\sqrt{1+c_{\pm}^2}},\quad
c_{\pm}=\frac{2\eta_2\pm \sqrt{4\eta_2^2 + 9 \eta_1^2}}{3\eta_1}.
\end{equation}
By making use of Eq. (\ref{hel}) we find that the minimum error
probability for identifying the state of the qubit $B$ is given by
\begin{widetext}
\begin{eqnarray}
P_E^{\prime}&=& \eta_{min} \left( 1 - \frac{1}{2}
 \frac{\eta_{max}}{\eta_{max}-\eta_{min} + \sqrt{(\eta_{max} - \eta_{min})^{2} + 2\eta_{min}\eta_{max}} } \right) ,
\label{pe1}
\end{eqnarray}
\end{widetext}
According to Eq. (\ref{opt}) it is reached with the help of the detection operators
\begin{equation}
\Pi_1^{\rm opt}= \left \{
\begin{array}{ll}
|\varphi_{-}\rangle_{BC} \langle \varphi_{-}|_{BC}
         \;\; & \mbox{if $\eta_1 \leq \frac{2}{5}$}\\
\\
         |u_1\rangle_{BC} \langle u_1|_{BC}+
|\varphi_{-}\rangle_{BC} \langle \varphi_{-}|_{BC}
                  \;\; & \mbox{otherwise,}
\end{array}
\right.
\nonumber
\end{equation}
and $\Pi_2^{\rm opt}= I_{BC} - \Pi_1^{\rm opt}$, where we have made use
of the identity
 $I_{BC}= |u_1\rangle \langle u_1| + |u_3\rangle \langle u_3|
+ \sum_{i=\pm} |\varphi_i\rangle \langle \varphi_i|.$
Clearly, the measurement that identifies the
state of the qubit $B$ with the smallest possible error is a
joint projection measurement on the qubits $B$ and $C$.

To close this section we briefly investigate the case that we can only perform a measurement on the qubit $B$ alone,
which means that our identification problem amounts to discriminating between the
density operator $\rho_{1}^{\prime \prime} = |0\rangle_B {}_{B}\langle 0|$ and
the uniformly mixed state $\rho_{2}^{\prime \prime} = {\rm Tr}_C \rho^{\prime}_2 = \frac{1}{2}I_B$.
For  minimum-error identification we consider the eigenvalues and
eigenstates of the operator
$\Lambda^{\prime \prime} = (\eta_2/2-\eta_1)\,|0\rangle_B {}_{B}\langle 0|
+ \eta_2/2 \,|1\rangle_B {}_{B}\langle 1|$, obtaining
in the standard way  the minimum error probability
\begin{equation}
\label{P_EB}
P_E^{\prime \prime}= \left \{
\begin{array}{ll}
\eta_1  \;\; & \mbox{if $\eta_1 \leq \frac{1}{3}$}\\
\frac{1}{2}(1-\eta_1)
         \;\; & \mbox{otherwise.}
\end{array}
\right.
\nonumber
\end{equation}
If $\eta_1\leq 1/3$ the error probability
is smallest when the qubit $B$ is always guessed to be in the
unknown state of the qubit $C$, while otherwise
a projection measurement characterized by the operators
$\Pi_1^{\rm opt}= |0\rangle_B {}_{B}\langle 0|$ and
$\Pi_2^{\rm opt}= |1\rangle_B {}_{B}\langle 1|$ has to be performed
in order to minimize the error.

As can be seen from Fig. 2, for $\eta_1\leq 2/5$ the optimum
two-qubit measurement leads to a substantial reduction of the minimum error
probability in comparison to the optimum single-qubit measurement.
The difference is largest for $\eta_1= 1/3$, where $P_E^{\prime}=0.22$
but $P_E^{\prime \prime} = 0.33$.
\begin{figure}[ht]
\begin{center}
\epsfig{file=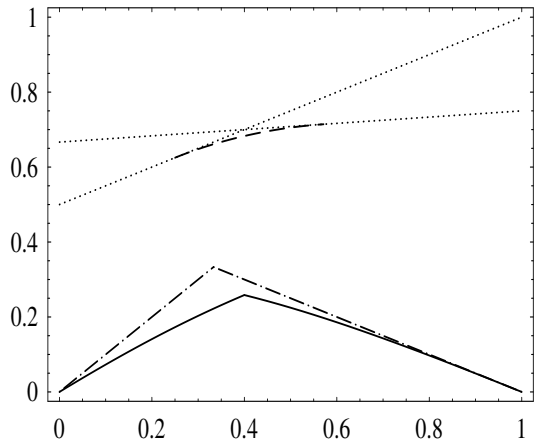,height=6cm,width=7cm}
\caption
{Failure and error probabilities for the various strategies of discriminating between one known and one unknown state vs. the prior probability $\eta_{1}$. The minimum error
probabilities $P_E^{\prime}$ resulting from
a joint measurement on the qubits $B$ and $C$ (full line) and
from a single-qubit measurement, $P_E^{\prime \prime}$ (dash-dotted line)
are compared to the failure probabilities for unambiguous identification
$Q_F^{\prime B}$, $Q_F^{\prime BC}$ (dotted lines)
and $Q_{F}^{\prime POVM}$ (dashed line).}
\end{center}
\end{figure}

\subsection{Optimum unambiguous discrimination between one known and one unknown state}

Finally we want to compare the minimum error probabilities with the
minimum failure probability in unambiguously identifying the qubit state.
For this purpose we again use the method described in \cite{HB3} and \cite{BFH}. By taking a reduction theorem \cite{raynal} into account, from Eqs. (\ref{sigma1k}) and (\ref{sigma2k}) it follows that the most general Ansatz for the detection operators can be written as
\begin{equation}
\label{op1}
\Pi_1=\alpha |\bar{u}_2\rangle \langle \bar{u}_2|,\qquad
\Pi_2=\beta |\bar{v}_2\rangle \langle \bar{v}_2| + |u_{3}\rangle \langle u_{3}| ,
\end{equation}
where $|\bar{v}_2\rangle = (|u_2\rangle - |\bar{u}_2\rangle)/\sqrt{2}$.
Here again the subscript $BC$ has been dropped.
Clearly, $\Pi_1\rho^{\prime}_2= \Pi_2\rho^{\prime}_1 =0$ as required for unambiguous discrimination.
As follows from Eq. (\ref{Qfail}), these detection operators yield the failure probability
\begin{equation}
Q_{fail}^{\prime} = 1- \frac{\eta_1}{4}\alpha - \frac{\eta_2}{6}(2 + \beta)
\label{Qfail2}
\end{equation}
which has to be minimized under the constraint that $\Pi_0 = I - \Pi_1- \Pi_2$ is
a positive operator. For $\Pi_{0}$ we obtain the expression
\begin{eqnarray}
\label{op3}
\Pi_0 &=& I - \alpha |\bar{u}_2\rangle \langle \bar{u}_2| - |u_3\rangle \langle u_3| \nonumber\\
{}&{}&-\frac{\beta}{2}(|u_2\rangle - |\bar{u}_2\rangle)(\langle u_2| - \langle \bar{u}_2|).
\end{eqnarray}
The eigenvalues of $\Pi_0$ are $\mu_1=1$, $\mu_2=0$ and
$\mu_{\pm}= (2-\beta-\alpha \pm \sqrt{\alpha^2+\beta^2})/2$,
and they all are non-negative provided that
$\beta \leq(2-2\alpha)/(2-\alpha)$. In order to minimize
$Q_{fail}$ while keeping $\Pi_0$ a positive operator we therefore choose
\begin{equation}
\label{op4}
\beta =\frac{2- 2\alpha} {2-\alpha}.
\end{equation}
Upon substituting  $\beta$  into Eq. (\ref{Qfail2}) and
determining the smallest value of the resulting function of $\alpha$, taking into account
that $0 \leq \alpha \leq 1$,  we find that the minimum failure probability is obtained when
\begin{equation}
\label{alpha}
\alpha= \left \{
\begin{array}{ll}
0  \;\; &  \mbox {if $ \;\;3\eta_1 \leq \eta_2 $}  \\
2\left(1-\sqrt{\frac{\eta_2}{3\eta_1}}\right)
         \;\; & \mbox{if $\;\;\eta_2\leq 3\eta_1 \leq 4\eta_2$}\\
 1  \;\; & \mbox{if $\;\;3\eta_1 \geq 4\eta_2$}.\\
\end{array}
\right.
\nonumber
\end{equation}
Using Eqs. (\ref{alpha}) and (\ref{op4}) in Eq. (\ref{op1}) yields the explicit expressions for
the optimum detection operators.

If $3\eta_1\leq \eta_2$, which implies that $\eta_1\leq 1/4$,
we have $\Pi_1^{\rm opt}=0$ and
$\Pi_2^{\rm opt} = |1\rangle_B\, {}_{B}\langle 1| \otimes I_{C}$, i. e. the
optimum measurement is a projection measurement on the qubit $B$ alone.
On the other hand, for $3\eta_1\geq 4\eta_2$, i.e.\  $\eta_1\geq 4/7$,
the optimum measurement is
a joint projection measurement on the qubits $B$ and $C$, where
$\Pi_1^{\rm opt}=|\bar{u}_2\rangle_{BC}\, {}_{BC}\langle \bar{u}_2|$ and
$\Pi_2^{\rm opt}= |1\rangle_B\, {}_{B}\langle 1| \otimes |1\rangle_C\, {}_{C}\langle 1|$.
The failure probability of these two von Neumann measurements is given by
\begin{eqnarray}
Q_F^{\prime B}&=& 1- \frac{1-\eta_1}{2} , \nonumber \\
Q_F^{\prime BC}&=& 1 - \frac{\eta_1}{4} - \frac{\eta_2}{3}=
\frac{2}{3} + \frac{\eta_1}{12}.
\end{eqnarray}
In the intermediate parameter region the optimum measurement is a
generalized measurement, yielding the failure probability
\begin{equation}
Q_F^{\prime POVM}=
 \frac{\eta_1}{6}+\frac{1+\sqrt{ 3\eta_1(1-\eta_1)}}{3}
        \;\;
\mbox{$ \left( \frac{1}{4}  \leq \eta_1 \leq \frac{4}{7}\right) $}.
\end{equation}

The benefit of performing the generalized measurement is only marginal, as evidenced by Fig. 2.
In fact, the reduction of the failure probability compared to the
best of the two types of von Neumann-measurements is largest for
$\eta_1 = 0.4$, where $Q_F^{\prime B}=Q_F^{\prime BC}= 0.7$, while the
generalized measurement yields the failure probability
$Q_F^{\prime POVM}=0.683$. In agreement with the general relation derived in
\cite{HB2}, the latter value is more than twice as large as the
minimum error probability $P_E^{\prime}$ which for
$\eta_1=0.4$ takes its maximum value 0.258.

A comparison of Figs. 1 and 2 reveals that, as expected, the minimum
probabilities of error and failure are
smaller when one reference state is known than in the case when both
reference qubits are in unknown states.

\section{Two-qubit program, $n$ copies of data state}

\subsection{Formulation of problem}
We now return to the situation in which we possess only one copy of each
of the two states we are trying to distinguish, but now we have $n>1$ copies
of the unknown state.  This means that we want a POVM that will distinguish
the two $n+2$ qubit states
\begin{eqnarray}
|\Psi_{1}\rangle & = & |\psi_{1}\rangle_{A}\otimes
|\psi_{1}\rangle_{1}\otimes \ldots |\psi_{1}\rangle_{n}\otimes|\psi_{2}\rangle_{C}\; , \nonumber \\
|\Psi_{2}\rangle & = & |\psi_{1}\rangle_{A} \otimes
|\psi_{2}\rangle_{1}\otimes \ldots
|\psi_{2}\rangle_{n}\otimes|\psi_{2}\rangle_{C} ,
\end{eqnarray}
that occur with {\it a priori} probabilities $\eta_{1}$ and $\eta_{2}$, respectively. 
Again we give two protocols, one for minimum error discrimination, and one for optimum 
unambiguous discrimination between the states.

To this end we now define the spaces and operators that we
will need.  Let $\Sigma$ be the space of symmetric states in
$\bigotimes^{n+1}\mathcal{H}$, where $\mathcal{H}$ is the two-dimensional
space for a single qubit. $\Sigma$ is an $n+2$ dimensional subspace.
$|\Psi_{1}\rangle$ is an element of $\Sigma\otimes\mathcal{H}=S_{1}$ and
$|\Psi_{2}\rangle$ is an element of $\mathcal{H}\otimes\Sigma = S_{2}$.
Their intersection, $S_{0}=S_{1}\cap S_{2}$, is the space of symmetric states in
$\bigotimes^{n+2}\mathcal{H}$. $S_{0}$ is a subspace of dimension $n+3$. Let
$K$ be the subspace of $\bigotimes^{n+2}\mathcal{H}$ generated by $S_{1}$ and
$S_{2}$. The dimension of $K$ is $3n+5$. Let $S_{3}$ be the
orthogonal complement of $S_{0}$ in $S_{1}$, let $S_{4}$ be the
orthogonal complement of $S_{0}$ in $S_{2}$, and let $L$ be the orthogonal
complement of $S_{0}$ in $K$. As was discussed in Section II, because we
do not know what $|\psi_{1}\rangle$ and $|\psi_{2}\rangle$ are, our problem
is to discriminate between the density matrices that result from averaging
$|\Psi_{1}\rangle\langle\Psi_{1}|$ and $|\Psi_{2}\rangle\langle\Psi_{2}|$
over $|\psi_{1}\rangle$ and $|\psi_{2}\rangle$.  This yields the two
density matrices
\begin{eqnarray}
\rho_{1} & = & \frac{1}{2n+4}P_{S_{1}} = \frac{1}{2n+4}P_{\Sigma}\otimes I , \nonumber \\
\rho_{2} & = & \frac{1}{2n+4}P_{S_{2}} = \frac{1}{2n+4}I\otimes P_{\Sigma} ,
\label{rho12n}
\end{eqnarray}
where $P_{S_{1}}$ and $P_{S_{2}}$ are the projections onto $S_{1}$ and
$S_{2}$, and $P_{\Sigma}$ and $I$ onto $\Sigma$ and $\mathcal{H}$, respectively.  Consequently, we reduced the problem to 
discriminating between the $2n+4$ dimensional spaces $S_{1}$ and $S_{2}$
in $K$, which is equivalent to discriminating between the $n+1$ dimensional 
subspaces $S_{3}$ and $S_{4}$ in the $2n+2$ dimensional space $L$.

We will now choose some bases in order to construct Jordan bases for
these subspaces. Jordan bases $\{ |p_{j}\rangle|j=0,\ldots N\}$ and 
$\{ |r_{j}\rangle|j=0,\ldots N\}$ for two $N+1$-dimensional subspaces, $S_{p}$ and
$S_{r}$, in general position, are orthonormal bases of their respective
subspaces ($\{ |p_{j}\rangle\}$ for $S_{p}$ and $\{ |r_{j}\rangle\}$ for $S_{r}$) that,
in addition, satisfy $\langle r_{j}|p_{k}\rangle = \delta_{jk}\cos\theta_{k}$.
The angles $\theta_{k}$ are called the Jordan angles. Now, let $|0\rangle$ and $|1\rangle$ be orthonormal basis vectors for $\mathcal{H}$. Further, let $|u_{j}^{(n+1)}\rangle$ ($j=0,\ldots n+1$) be the unique unit vector in the symmetric subspace of $n+1$ qubits, $\Sigma$, which is the sum of $n+1$-tuples with $j$ ones and $n-j+1$ zeros,
\begin{eqnarray}
|u_{0}^{(n+1)}\rangle &=& |0\ldots0\rangle \ , \nonumber \\
|u_{1}^{(n+1)}\rangle &=& \frac{|0\ldots 01\rangle + |0 \ldots 10\rangle + \ldots + |10 \ldots 0\rangle}{\sqrt{n+1}} \ , \nonumber \\
{}&\vdots &  \nonumber \\
|u_{n+1}^{(n+1)}\rangle &=& |11\ldots 1\rangle \ .
\label{Sigmabasis}
\end{eqnarray}

Then the structure of the two density operators in (\ref{rho12n}), in particular the  decomposition on the right hand side, suggests that we consider 
$|e_{j,\alpha}\rangle= |u_{j}^{(n+1)}\rangle\otimes|\alpha\rangle$ and
$|f_{j,\alpha}\rangle= |\alpha\rangle\otimes|u_{j}^{(n+1)}\rangle$ where
$\alpha = 0,1$ and $0 \leq j \leq n+1$. These vectors form
orthonormal bases for $S_{1}$ and $S_{2}$, respectively.  Let
$|u_{j}^{(n+2)}\rangle$ ($j=0,\ldots n+2$) be the unique unit vector in the symmetric subspace of $n+2$ qubits, $S_{0}$, which is the sum of $n+2$-tuples with exactly $j$ ones and $n+2-j$ zeros.  
This vector can be expressed in terms of either the $S_{0}$ or $S_{1}$
basis, since it is in both spaces. A direct calculation shows that 
\begin{equation}
|u_{j}^{(n+2)}\rangle=\sqrt{\frac{n+2-j}{n+2}}|e_{j,0}\rangle 
+ \sqrt{\frac{j}{n+2}}|e_{j-1,1}\rangle , 
\end{equation}
and
\begin{equation}
|u_{j}^{(n+2)}\rangle= \sqrt{\frac{n+2-j}{n+2}}|f_{j,0}\rangle +
\sqrt{\frac{j}{n+2})}|f_{j-1,1}\rangle 
\end{equation}
for $0 \leq j \leq n+2$. In particular, $|u_{0}^{(n+2)}\rangle =|0,0,...,0\rangle$ 
and $|u_{n+2}^{(n+2)}\rangle) = |1,1,...,1\rangle$.

We now introduce the vectors 
\begin{equation}
|g_{j}\rangle = \sqrt{\frac{j}{n+2}}|e_{j,0}\rangle - \sqrt{\frac{n+2-j}{n+2}}|e_{j-1,1}\rangle , 
\end{equation}
and 
\begin{equation}
|h_{j}\rangle = \sqrt{\frac{j}{n+2}}|f_{j,o}\rangle - \sqrt{\frac{n+2-j}{n+2}}|f_{j-1,1}\rangle , 
\end{equation}
for $1 \leq j\leq n+1$.
The $|g_{j}\rangle$'s and $|h_{j}\rangle$'s form orthonormal bases
for $S_{3}$ and $S_{4}$. Each vector on the right-hand sides
of the above expressions has exactly $j$ ones. Therefore, if $j\neq k$
\begin{equation}
\langle g_{j}|h_{k}\rangle= 0 , 
\end{equation}
and $\{ |g_{j}\rangle\}$ and $\{ |h_{j}\rangle\}$ form Jordan bases for
$S_{3}$ and $S_{4}$. Let $T_{j}$ be the two dimensional vector space spanned by the nonorthogonal but linearly independent vectors $|g_{j}\rangle$ and $|h_{j}\rangle$. The $T_{j}$ form a
decomposition of $L$ into $n+1$ mutually perpendicular two
dimensional subspaces.  A calculation shows that 
\begin{eqnarray}
\langle f_{j,0}|e_{j,0}\rangle= \frac{n+1-j}{n+1}, \nonumber \\
\langle f_{j-1,1}|e_{j-1,1}\rangle = \frac{j-1}{n+1}, 
\end{eqnarray} 
and
\begin{equation} 
\langle e_{j-1,1}|f_{j.0}\rangle = \langle
f_{j-1,1}|e_{j,0}\rangle = \frac {(j(n+2-j))^{1/2}}{n+1} .
\end{equation}
Therefore,
\begin{equation}
\langle h_{j}|g_{j}\rangle= -\frac{1}{n+1} , 
\end{equation}
and the Jordan angles are all the same. 
The two density operators that we wish to
distinguish can now be expressed as 
\begin{eqnarray}
\rho_{1}& = & \frac{1}{2(n+2)}[P_{S_{0}} +
\sum_{i=1}^{n+1} |g_{i}\rangle\langle g_{i}|]
\nonumber \\ 
\rho_{2} & = & \frac {1}{2(n+2)}[P_{S_{0}} +
\sum_{i=1}^{n+1}|h_{i}\rangle\langle h_{i}] ,
\label{rho12ndiagonal}
\end{eqnarray}
where
\begin{equation}
P_{S_{0}}=\sum_{j=0}^{n+2}|u_{j}^{(n+2)}\rangle \langle u_{j}^{(n+2)}|
\label{S0projector}
\end{equation}
is the projection onto $S_{0}$.

\subsection{Minimum error discrimination strategy}

For minimum-error identification we have to determine the eigenvalues and eigenstates of the 
operator $\Lambda = \eta_{2}\rho_{2}-\eta_{1}\rho_{1}$ which, using Eq. (\ref{rho12ndiagonal}), can be written as
\begin{eqnarray}
\Lambda & = &  \frac{1}{2n+4}[(\eta_{2}- \eta_{1})P_{S_{0}}+
\eta_{2}\sum|h_{i}\rangle\langle h_{i}| \nonumber \\
 & & -\eta_{1}\sum|g_{i}\rangle\langle g_{i}|] . 
\end{eqnarray}
$\Lambda$ is diagonal in $S_{0}$ and it is straightforward to carry out the diagonalization in each of the two-dimensional subspaces spanned by $|g_{i}\rangle$ and $|h_{i}\rangle$, yielding the  spectral representation,
\begin{equation}
\label{Lambda12n}
\Lambda = \sum_{j=0}^{n+2}\lambda_{0}|u_j\rangle \langle u_j| + \sum_{i=1}^{n+1} \left(\lambda_{+} |\varphi_{i+}\rangle \langle \varphi_{i+}| + \lambda_{-}|\varphi_{i-}\rangle \langle \varphi_{i-}|\right) ,  
\end{equation}
where
\begin{equation}
\lambda_{0} =  \frac{\eta_{2} - \eta_{1}}{2n+4} \ ,
\end{equation}
and
\begin{eqnarray}
\lambda_{\pm} &=& \frac{1}{4(n+2)}\left( \eta_{2} - \eta_{1}\right. \nonumber \\
 {}&&\left. \pm \sqrt{\left(\eta_{2} - \eta_{1}\right)^2 +\frac{4n(n+2)}{(n+1)^{2}} \eta_{1} \eta_{2}}\  \right) \ .
\end{eqnarray}
The eigenvalue associated with $S_{0}$, $\lambda_{0}$, has a degeneracy $(n+3)$ and the eigenvalues associated with $S_{3}$ and $S_{4}$, $\lambda_{\pm}$, have a degeneracy $(n+1)$ each. Furthermore, in the nonorthogonal basis of the two-dimensional subspace $T_{i}$ given by $|g_{i}\rangle$ and $|h_{i}\rangle$, 
\begin{equation}
 \label{phipm}
|\varphi_{i\pm}\rangle=\frac{|g_i\rangle-c_{\pm}|h_i\rangle}{\sqrt{1+c_{\pm}^2-2c_{\pm}/(n+1)}},
\end{equation}
where
\begin{equation}
c_{\pm}=(n+1)\left[1 -\frac{\frac{\eta_{2}}{(n+1)^{2}} - \eta_{1}}{(2n+4)\lambda_{\pm}}\right].
\end{equation}
We find that $\lambda_{-}$ is unconditionally negative, $\lambda_{0}$ is negative if $\eta_{2} < \eta_{1}$ and positive otherwise, and $\lambda_{+}$ is unconditionally positive.  By making use of Eq. (\ref{hel}) we find that the minimum error 
probability for identifying the state of the data qubits is given by
\begin{widetext}
\begin{eqnarray}
P_{E}= \eta_{min}\left[1 - \frac{n}{n+1} \frac{\eta_{max}}{\eta_{max}-\eta_{min}+\sqrt{(\eta_{max}-\eta_{min})^{2} + \frac{4n(n+2)}{(n+1)^{2}}\eta_{min}\eta_{max}}} \right] \ ,
\end{eqnarray}
\end{widetext}
where $\eta_{min}$ ($\eta_{max}$) is the smaller (larger) of $\{\eta_{1},\eta_{2}\}$. According to Eq. (\ref{opt}), the minimum error probability is reached with the help of the optimum detection operators 
\begin{equation}
\Pi_1^{\rm opt}= \left \{
\begin{array}{ll}
\sum_{i=1}^{n+1}|\varphi_{i-}\rangle \langle \varphi_{i-}|
         \;\; & \mbox{if $\eta_1 \leq \frac{1}{2}$} \\
\\
         P_{S_{0}} +  
\sum_{i=1}^{n+1}|\varphi_{i-}\rangle \langle \varphi_{i-}|
                  \;\; & \mbox{if $\eta_1 > \frac{1}{2}$} 
\end{array} \ ,
\right. 
\nonumber
\end{equation}
which is the projection onto the strictly negative eigensapce of $\Lambda$,
and $\Pi_2^{\rm opt}= I_{K} - \Pi_1^{\rm opt}$, where the identity
 $I_{K}= P_{S_{0}} + \sum_{i=1}^{n+1}( |\varphi_{i+}\rangle \langle \varphi_{i+}| + |\varphi_{i+}\rangle \langle \varphi_{i+}|)$. $P_{S_{0}}$ is given in Eq. (\ref{S0projector}).   
Clearly, the measurement that identifies the state of the data qubits with the smallest possible error is a joint projection measurement on all of the qubits. It should be noted that for $n=1$ the formulas in this Section reduce to those of Sec. III.A whereas for $n \rightarrow  \infty$ we have that $P_{E} \rightarrow \eta_{min}/2 \leq 1/4$.

\subsection{The optimal universal unambiguous bound}
We now want to consider the unambiguous
discrimination between the subspaces $S_{1}$ and $S_{2}$ in $K$, or
equivalently between $S_{3}$ and $S_{4}$ in $L$. Let $S_{i}^{\perp}$
be the orthogonal complement of $S_{i}$ in $K$. $S_{i}^{\perp}$ is
equal to the orthogonal complement of $S_{i+2}$ in $L$.
$S_{i}^{\perp}$ is an $n+1$ dimensional subspace. The POVM which
unambiguously distinguishes between $S_{1}$ and $S_{2}$ has the form
$\Pi_{1}=\alpha P_{S_{2}^{\perp}}$, and $\Pi_{2}=\beta P_{S_{1}^{\perp}}$,
where the $P$'s are orthogonal projections onto $S_{1}^{\perp}$ or 
$S_{2}^{\perp}$, and the $\alpha$ and $\beta$ are positive real numbers
between zero and one, which are so chosen that $\Pi_{1}$, $\Pi_{2}$,
and $\Pi_{0}=I- \Pi_{1}- \Pi_{2}$ are the elements of a POVM on $K$.

Let us define $|g_{i}^{\perp}\rangle$ in $S_{1}^{\perp}$, and
$|h_{i}^{\perp}\rangle$ in $S_{2}^{\perp}$ by the formulas
\begin{eqnarray}
|h_{i}\rangle & = & -\frac{1}{n+1}|g_{i}\rangle+
\frac{\sqrt{n(n+2)}}{n+1}|g_{i}^{\perp}\rangle \nonumber \\
|g_{i}\rangle & = -& \frac{1}{n+1}|h_{i}\rangle+ \frac
{\sqrt{n(n+2)}}{n+1}|h_{i}^{\perp}\rangle , 
\end{eqnarray}
on $T_{i}$, and we have that
\begin{eqnarray}
P_{S_{1}^{\perp}}  =  \sum_{i=1}^{n+1} |g_{i}^{\perp}\rangle
\langle g_{i}^{\perp}| ,\ \ \ \
P_{S_{2}^{\perp}}  =  \sum_{i=1}^{n+1}  |h_{i}^{\perp}\rangle
\langle h_{i}^{\perp}| . 
\end{eqnarray}
The $\alpha$ and $\beta$ can now be chosen so that $\Pi_{0}$
restricted to each $T_{i}$ is non-negative. The matrix which
represents $\Pi_{0}$ on $T_{i}$, in the basis
$\{|g_{i}\rangle,|g_{i}^{\perp}\rangle\}$, is
\begin{equation}
\left(
  \begin{array}{cc}
    1-\alpha \frac{n^{2}+2n}{(n+1)^{2}} &  -\alpha
\frac{\sqrt{n(n+n)}}{(n+1)^{2}} \\
    -\alpha\frac{\sqrt{n(n+n)}}{(n+1)^{2}} &
     1- \beta - \frac{\alpha}{(n+1)^{2}} \\
  \end{array}
\right) .
\end{equation}
This matrix must be positive.  Therefore, 
$\Pi_{1}$, $\Pi_{2}$, and $\Pi_{0}$ form a POVM if and only if

\begin{equation}
1- \alpha - \beta - \alpha \beta \frac{n^{2}+2n}{(n+1)^{2}} \geq 0 ,
\label{poscond}
\end{equation}
provided that
\begin{equation}
0 \leq \alpha, \beta \leq 1 .
\label{probcond}
\end{equation}
We also have that 
\begin{eqnarray}
{\rm Tr}(\Pi_{1}\rho_{1}) = \frac{\alpha n}{2n+2} , \ \ \
{\rm Tr}(\Pi_{2}\rho_{2}) = \frac{\beta n}{2n+2} .
\end{eqnarray}

If we are given $\rho_{1}$ with probability $\eta_{1}$ and
$\rho_{2}$ with probability $\eta_{2}$ then the probability
that this POVM successfully distinguishes $S_{1}$ from $S_{2}$ is
\begin{equation}
P_{succ}=\frac{n}{2n+2}(\eta_{1}\alpha + \eta_{2}\beta) ,
\end{equation}
if the $\alpha$ and $\beta$ satisfy the constraints above. If we set
\begin{equation}
\label{c2}
\beta = \frac{1- \alpha}{1- \frac{n^{2}+ 2n}{(n+1)^{2}}\alpha} 
\end{equation}
in the above formula, which is the maximum allowed by (\ref{poscond}), and differentiate, we find that $P_{succ}$ has a maximum value when
\begin{equation}
\label{c1}
\alpha= \left \{
\begin{array}{ll}
0   &  \mbox {if $\eta_1 \leq \frac{1}{1+(n+1)^{2}}$}  \\
\frac{n+1}{n^{2}+n}\left(n+1-\sqrt{\frac{\eta_2}{\eta_1}}\right)
    & \mbox{if $\frac{1}{1+(n+1)^{2}} \leq \eta_1 \leq \frac{(n+1)^{2}}{1+(n+1)^{2}}$}\\
1   & \mbox{if $\eta_1 \geq \frac{(n+1)^{2}}{1+(n+1)^{2}}$}.\\
\end{array}
\right.
\nonumber
\end{equation}
The maximum value of $P_{succ}$ is
\begin{equation}
P_{max}=\frac{1}{n+2}[\frac{n+1}{2}- (\eta_{1}\eta_{2})^{1/2}] ,
\label{Psuccmax}
\end{equation}
provided $\frac{1}{1+(n+1)^{2}} \leq \eta_{1} \leq \frac{(n+1)^{2}}{1+(n+1)^{2}}$, using the center line in (\ref{c1}). This clearly can be obtained by a POVM only. If $\eta_{1}$ is to the left of this interval (first line in (\ref{c1})) the optimum measurement is the projection $P_{S_{1}^\perp}$, which unambiguously identifies
$\rho_{2}$ 
with a success probability $P_{succ}^{(2)} = \eta_{2}n/(2n+2)$. If $\eta_{1}$ is to the right of this interval (last line in (\ref{c1})) the optimum measurement is the projection
$P_{S_{2}^{\perp}}$ which unambiguously identifies $\rho_{1}$ 
with a success probability $P_{succ}^{(1)}=\eta_{1}n/(2n+2)$. It should be noted that the optimum failure probability is given as $Q_{F}=1-P_{succ}$. For $n=1$ these expressions reproduce the corresponding ones in Sec. III.B. For $\eta_{1}=\eta_{2}=1/2$, when their difference is the largest, $Q_{F}^{POVM}=1-P_{max}=\frac{n+4}{2n+4}$ and $Q_{F}^{(1,2)}=1-P_{succ}^{(1,2)}=\frac{3n+4}{4n+4}$, as a function of $n$. For $n \rightarrow \infty$ we have $Q_{F}^{POVM} \rightarrow 1/2$ and $Q_{F}^{(1,2)}\rightarrow 3/4$, so the POVM outperforms the projective measurements quite significantly.  Furthermore, $P_{E} \leq Q_{F}^{POVM}/2$ always holds, as it should.

If we use the universal POVM $\Pi_{1}$, $\Pi_{2}$, and $\Pi_{0}$ to
unambiguously discriminate between the states $|\Psi_{1}\rangle$ and
$|\Psi_{2}\rangle$ without averaging over them, we find that $\tilde{P}_{succ}$ the probability of 
success is
\begin{eqnarray}
\tilde{P}_{succ} & = & \eta_{1}\alpha \langle \Psi_{1}| \Pi_{1}|\Psi_{1}\rangle
 + \eta_{2}\beta \langle\Psi_{2}|\Pi_{2}|\Psi_{2}\rangle \nonumber \\
 & = & \frac{n}{n+1}(\eta_{1}\alpha + \eta_{2}\beta)(1-
|\langle\psi_{1}|\psi_{2}\rangle|^{2}) ,
\end{eqnarray}
where the $\alpha$ and $\beta$ satisfy the same constraints as above. 
Inserting their optimal values from Eqs.\ (\ref{c1}) and (\ref{c2}) we find
\begin{equation}
\tilde{P}_{opt}=\frac{1-|\langle\psi_{1}|\psi_{2}\rangle|^{2}}{n+2} [n+1-2(\eta_{1}\eta_{2})^{1/2}] ,
\label{popt3} 
\end{equation}
with the same restrictions on
$\eta_{1}$ as in the previous paragraph. If we average the overlap term
over all possible choices of the $|\psi_{i}\rangle$'s we can replace
it with its average value of 1/2, and we recover (\ref{Psuccmax}).

As expected, the optimal success probability is an increasing function of $n$.  The more copies of
the unknown qubit we possess, the greater our chance of identifying it.

The $n\rightarrow\infty$ limit of $\tilde{P}_{opt}$ can also be achieved by a
different strategy than the one we are employing here.  With a very
large number of copies of the unknown qubit, we could employ state
reconstruction techniques to find out its state \cite{buzek}. For example, suppose
we have determined the state of the unknown qubit to be
$|\psi_{0}\rangle$.  While we know what this state is, we do not know if
it is equal to the first or the second program state, $|\psi_{1}\rangle$ or $|\psi_{2}\rangle$.  We therefore project each of the two program states onto the state
orthogonal to the reconstructed state. That is, we take the program qubit that we know is in the state $|\psi_{i}\rangle$ ($i=1$ or $2$) and
measure the projection operator $P_{0\perp}=|\psi_{0}^{\perp}\rangle
\langle\psi_{0}^{\perp}|$.  If $|\psi_{0}\rangle = |\psi_{1}\rangle$, which is given with the {\it a priori} probability $\eta_{1}$, then
this measurement succeeds (gives $1$) with a probability of
$\langle\psi_{1}^{\perp}|\psi_{2}\rangle |^{2}$, and if $|\psi_{0}\rangle
= |\psi_{2}\rangle$, which is given with the {\it a priori} probability $\eta_{2}$, it succeeds with the same probability.  Therefore, the total probability of
success is just $(\eta_{1}+\eta_{2})|\langle\psi_{1}^{\perp}|\psi_{2}\rangle |^{2}$,
which is the same as the $n\rightarrow\infty$ limit of Eq.\ (\ref{popt3}).
While these strategies give the same result for an infinite number of
copies, there is a difference between them for $n$ finite.  The strategy
that led to Eq.\ (\ref{popt3}), will never produce an erroneous result,
while the strategy based on state reconstruction can.  If our determination
of the state of the unknown qubit is not exact, which will, in general, be
the case for finite $n$, then $|\psi_{0}^{\perp}\rangle$ will not be
orthogonal to either $|\psi_{1}\rangle$ or $|\psi_{2}\rangle$, and this
will lead to errors.

\section{One-qubit program, $n$ copies of data state}

In this section we return to the case when one of the states to be determined is known and the other is unknown. However, in contrast to the treatment that was presented in Section IV, we are now provided $n$ copies of the states to be determined. In other words, we now have $n$ data registers, $B_{1},\ldots,B_{n}$, and in each one we either have a copy of the known state $|0\rangle$ or a copy of the unknown state $|\psi\rangle$ stored in the program register $C$.  Our task is then to 
decide whether the qubit $B_{i}$ is in the known state $|0\rangle_{B_{i}}$, 
or whether it is in the unknown state of the program qubit stored in register $C$, for all $i$.

Thus, we assume that we have a system of $n+1$ qubits, labeled $B_{1},\ldots,B_{n}$, and $C$ where $C$ is the program qubit and $B_{1},\ldots,B_{n}$ are the data qubits. Qubit $C$ is always prepared in the state $|\psi\rangle$. Qubits $B_{i}$, the data qubits, are guaranteed to be all prepared in either $|0\rangle$ or all in $|\psi\rangle$, but we do not know which of these two alternatives occurs. The prior probabilities of these two alternatives are $\eta_1$ and $\eta_2=1-\eta_1$, respectively.  Our task is then to find the optimal measurement (POVM) that will distinguish
the two $n+1$ qubit states,
\begin{eqnarray}
|\Psi_{1}\rangle  &=& |0\rangle_{1}    \otimes \ldots |0\rangle_{n} \otimes | \psi \rangle_{C} \ , \nonumber \\
|\Psi_{2}\rangle &=& |\psi \rangle_{1} \otimes \ldots |\psi\rangle_{n} \otimes |\psi\rangle_{C} \ ,
\end{eqnarray}
where we dropped the subscript $B$ for the data registers as it leads to no confusion.

If the state $|\psi\rangle$ is completely unknown, we have to find the best measurement strategy that is optimal on average. Thus, we have to take the average of the input with respect to all possible qubit states. The identification problem is then equivalent to distinguishing between two mixed states, given by the density operators
\begin{eqnarray} 
\label{rh1n}
\rho_1^{\prime} &=& \left \{|\Psi_ {1}\rangle \langle \Psi_ {1}|\right \}_{\rm av} \nonumber \\ {}&=& \left \{ |0\rangle^{\otimes (n)}|\psi\rangle \langle 0|^{\otimes (n) } \langle\psi|\right \}_{\rm av}, \\
\label{rh2n}
\rho_2 ^{\prime}&=& \left \{|\Psi_{2}\rangle \langle \Psi_{2}|\right \}_{\rm av} \nonumber \\ {} &=& \left \{|\psi\rangle^{\otimes (n+1)} \langle \psi|^{\otimes (n+1)}\right \}_{\rm av},
\end{eqnarray}
that occur with the prior probabilities $\eta_1$ and $\eta_2$, respectively. The unknown qubit state can be again represented using the Bloch parametrization as $|\psi\rangle = \cos (\theta/2) |0\rangle + e^{i\phi} \sin (\theta/2) |1\rangle$, with $|0\rangle$ and $|1\rangle$ denoting an arbitrary set of orthonormal basis states. Here $\theta$ and $\phi$ are the polar and azimuthal angle on the Bloch sphere. After performing the averaging with respect to all possible values of $\theta$ and $\phi$ we arrive at 
\begin{eqnarray}
\label{rho1an}
\rho_1^{\prime} &=& \frac{1}{2} |0\rangle^{\otimes (n)} \langle 0|^{\otimes (n)} \otimes I_{C} \ ,\\ 
\label{rho2an}
\rho_2 ^{\prime}&=& \frac{1}{(n+2)} P_{B_{1} \ldots B_{n}C}^{sym} \ . 
\end{eqnarray}   
where $P_{B_{1} \ldots B_{n}C}^{sym}$ is the projector to the $n+2$ dimensional symmetric subspace of the corresponding $n+1$ qubits, $B_{1} \ldots B_{n}C$, and
$I_{C} =  |0\rangle_{C} {}_{C}\langle 0| + |1\rangle_{C} {}_{C}\langle 1|$. Eqs. (\ref{rho1an}) and (\ref{rho2an}) reduce the problem of the programmable state discriminator with single-qubit program and $n$ copies of the data state to the problem of discriminating between these two mixed states. 

At this point it is useful to introduce the following basis for the the $n+2$-dimensional symmetric subspace of the $n+1$ qubits
\begin{eqnarray}
|u_{1}\rangle &=& |0\ldots0\rangle \ , \nonumber \\
|u_{2}\rangle &=& \frac{|0\ldots 01\rangle + |0 \ldots 10\rangle + \ldots + |10 \ldots 0\rangle}{\sqrt{n+1}} \ , \nonumber \\
{}&\vdots &  \nonumber \\
|u_{n+2}\rangle &=& |11\ldots 1\rangle \ .
\label{symbasis}
\end{eqnarray}
We also introduce
\begin{equation}
|v_{2}\rangle = |0 \ldots 01\rangle \ .
\label{vtwo}
\end{equation}
The two nonorthogonal but linearly independent vectors, $|u_{2}\rangle$ and $|v_{2}\rangle$, span a two-dimensional subspace of the entire Hilbert space. It will prove useful later on to define two other vectors in this subspace as
\begin{equation}
\label{vtwobar}
|\bar{v}_{2}\rangle = \frac{|0 \ldots 10\rangle + \ldots + |10 \ldots 0\rangle}{\sqrt{n}} \ .
\end{equation}
and
\begin{eqnarray}
\label{utwobar}
|\bar{u}_{2}\rangle = \frac{1}{\sqrt{n+1}} |\bar{v}_{2}\rangle - \sqrt{\frac{n}{n+1}}|v_{2}\rangle \ ,
\end{eqnarray}
The two sets, $\{|v_{2}\rangle,|\bar{v}_{2}\rangle\}$ and $\{|u_{2}\rangle,|\bar{u}_{2}\rangle\}$, each form an orthonormal basis in the two-dimensional subspace. We then have to distinguish between the density operators $\rho^{\prime}_1$ and $\rho^{\prime}_2$ given by Eqs. (\ref{rho1an}) and (\ref{rho2an}) that refer to the first and second alternative, respectively, and that occur with the prior probabilities $\eta_1$ and $\eta_2$.
After reexpressing $\rho^{\prime}_1$ in terms of the basis states 
$|u_i\rangle$ ($i=1,\ldots n+2$) and $|v_{2}\rangle$, defined by Eqs. (\ref{symbasis}) and (\ref{vtwo}), respectively, the density operators to be discriminated read
\begin{eqnarray} 
\label{sigma1l1}
\rho^{\prime}_1 & =& \frac{1}{2}\left(|u_1\rangle \langle u_1| + |v_{2}\rangle \langle v_{2}| \right), \\
\label{sigma2l1}
\rho^{\prime}_2 & = & 
 \frac{1}{n+2} \left( \sum_{i=1}^{n+2}|u_i\rangle \langle u_i| \right) \ .
\end{eqnarray}
The subsequent treatment proceeds along exactly the same lines that we followed in  the previous sections.

\subsection{Minimum-error discrimination}
   
For identifying the data state with minimum error we have to determine the eigenvalues and eigenstates of the 
operator  $\Lambda^{\prime} = \eta_2 \rho^{\prime}_2 - \eta_1 \rho^{\prime}_1$ where $\rho^{\prime}_1$ and $\rho^{\prime}_2$ are given by Eqs. (\ref{rho1an}) and (\ref{rho2an}). From the explicit expression of these two density operators it is clear that $\Lambda^{\prime}$ is diagonal except in the two-dimensional subspace spanned by $|v_{2}\rangle$ and $|\bar{v}_{2}\rangle$.  It is straightforward to carry out the diagonalization in this subspace yielding the  spectral representation, 
\begin{equation}
 \label{Lambda31}
\Lambda^{\prime} = \lambda_{1}|u_1\rangle \langle u_1| + \sum_{i=-}^{+} \lambda_i |\varphi_i\rangle \langle \varphi_i| + \sum_{j=3}^{n+2}\lambda_{j}|u_j\rangle \langle u_j| ,  
\end{equation}
where
\begin{equation}
\lambda_{1} = \left( \frac{\eta_2}{n+2}- \frac {\eta_1}{2} \right) \ ,
\end{equation}
\begin{eqnarray}
\lambda_{\pm} &=& \frac{1}{2}\left( \frac{\eta_2}{n+2} - \frac{\eta_1}{2}\right. \nonumber \\
 {}&&\left. \pm \sqrt{\left(\frac{\eta_2}{n+2} - \frac{\eta_1}{2}\right)^2 +\frac{2 \eta_{1} \eta_{2} n}{(n+1)(n+2)}}\  \right),
\end{eqnarray}
and
\begin{equation}
\lambda_{j} = \frac{\eta_{2}}{n+2} \ ,
\end{equation}
for $j \geq 3$. Furthermore
\begin{equation}
 \label{phi31}
|\varphi_{\pm}\rangle=\frac{|v_2\rangle-c_{\pm}|\bar{v}_2\rangle}{\sqrt{1+c_{\pm}^2}},
\end{equation}
where
\begin{equation}
c_{\pm}=\frac{n+1}{2\sqrt{n}} \frac{\eta_{2}^{\prime} +\frac{n-1}{2(n+1)}\eta_{1} \pm \sqrt{(\eta_{2}^{\prime} - \eta_{1})^{2}+\frac{4n}{n+1}\eta_{2}^{\prime}\eta_{1}}}{\eta_1}.
\end{equation}
Here we introduced $\eta_{2}^{\prime}=2\eta_{2}/(n+2)$ which is the weight of $\rho_{2}^{\prime}$ in the intersection of the supports of the two density operators to be discriminated. We find that $\lambda_{-}$ is unconditionally negative, $\lambda_{1}$ is negative if $\eta_{2}/(n+2) < \eta_{1}/2$ and positive otherwise, and $\lambda_{+},\lambda_{j} \geq 0$ for $j \geq 3$.  By making use of Eq. (\ref{hel}) we find that the minimum error 
probability for identifying the state of the data qubits is given by
\begin{widetext}
\begin{eqnarray}
P_{E}^{\prime}= \eta_{min}\left[1 - \frac{n}{n+1} \frac{\eta_{max}}{\eta_{max}-\eta_{min}+\sqrt{(\eta_{max}-\eta_{min})^{2} + \frac{4n\eta_{min}\eta_{max}}{n+1}}} \right] \ ,
\end{eqnarray}
\end{widetext}
where we introduced $\eta_{min}$ ($\eta_{max}$) as the smaller (larger) of $\{\eta_{1},2\eta_{2}/(n+2)\}$. According to Eq. (\ref{opt}), the minimum error probability is reached with the help of the detection operators 
\begin{equation}
\Pi_1^{\rm opt}= \left \{
\begin{array}{ll}
|\varphi_{-}\rangle \langle \varphi_{-}|
         \;\; & \mbox{if $\eta_1 \leq \frac{2}{n+4}$} \\
\\
         |u_1\rangle \langle u_1|+  
|\varphi_{-}\rangle \langle \varphi_{-}|
                  \;\; & \mbox{if $\eta_1 > \frac{2}{n+4}$} 
\end{array} \ ,
\right. 
\nonumber
\end{equation}
and $\Pi_2^{\rm opt}= I - \Pi_1^{\rm opt}$, where we have to use the identity
 $I= |u_{1}\rangle \langle u_{1}| + \sum_{i=3}^{n+2} |u_i\rangle \langle u_i| + \sum_{i=-}^{+} |\varphi_i\rangle \langle \varphi_i|$.    
Clearly, the measurement that identifies the 
state of the data qubits with the smallest possible error is a 
joint projection measurement on the qubits $B_{i}$ (for $i=1,\ldots,n$) and $C$. It should be noted that for $n=1$ the formulas in this Section reduce to those of Sec. IV.A whereas for $n \rightarrow  \infty$ we have that $P_{E} \rightarrow 0$ since in this latter case the mixed states that we are trying to distinguish become essentially orthogonal. 
The vanishing of the error probability for $n\rightarrow \infty$ is in accordance with
the fact that the data state can be in principle exactly determined by tomographic methods, without any joint measurement, provided that an infinite number of copies is available. After the data state has been determined, it is of 
course possible to tell without error whether it is equal to the state $|0\rangle$ or not.

\subsection{Unambiguous discrimination}

Finally we want to determine the minimum failure probability for the unambiguous discrimination between the states given by (\ref{sigma1l1}) and  (\ref{sigma2l1}). 
For this purpose we again use the method described in \cite{HB3} and \cite{BFH}. Taking one of the reduction theorems derived in  \cite{raynal} into account, the most general Ansatz for the detection operators can be written as
\begin{equation}
\label{op11}
\Pi_1=\alpha |\bar{u}_2\rangle \langle \bar{u}_2|,\qquad
\Pi_2=\beta |\bar{v}_2\rangle \langle \bar{v}_2| + \sum_{i=3}^{n+2}|u_{i}\rangle \langle u_{i}| ,
\end{equation}
where $|\bar{v}_2\rangle$ and $|\bar{u}_{2}\rangle$ were given in (\ref{vtwobar}) and (\ref{utwobar}), respectively.
Clearly, $\Pi_1\rho^{\prime}_2= \Pi_2\rho^{\prime}_1 =0$ as required for unambiguous discrimination.
As follows from Eq. (\ref{Qfail}), these detection operators yield the failure probability
\begin{equation} 
Q_{fail}^{\prime} = 1- \frac{\eta_{1}\alpha n}{2(n+1)} - \frac{\eta_{2} n}{n+2} - \frac{\eta_{2}\beta n}{(n+1)(n+2)}
\label{Qfail21}
\end{equation} 
which again has to be minimized under the constraint that $\Pi_0 = I - \Pi_1- \Pi_2$ is 
a positive operator, in complete analogy to our procedure in Sec. IV A. For $\Pi_{0}$ we obtain the expression
\begin{eqnarray}
\label{op31}
\Pi_0 &=& |u_{1}\rangle \langle u_{1}| - \alpha |\bar{u}_{2}\rangle \langle \bar{u}_{2}| + |v_2\rangle \langle v_2| \nonumber\\
{}&{}&-\beta|\bar{v}_{2}\rangle \langle \bar{v}_{2}| + |\bar{v}_{2}\rangle \langle\bar{v}_{2}| \ .
\end{eqnarray}
The eigenvalues of $\Pi_0$ are $\mu_{1}=1$, $\mu_{i}=0$ for $3 \leq i \leq n+2$, and 
$\mu_{\pm}= (2-\beta-\alpha \pm \sqrt{(\alpha+\beta)^2-4\alpha\beta n/(n+1)})/2$. They all are non-negative provided that  
$\beta \leq (n+1)(1-\alpha)/(n+1-n\alpha)$. In order to minimize  
$Q_{fail}$ while keeping $\Pi_0$ a positive operator we therefore choose 
\begin{equation}
\label{op41}
\beta =\frac{(n+1)(1 - \alpha)} {n+1-n\alpha}.
\end{equation}
Upon substituting this expression into Eq. (\ref{Qfail21}) the failure probability becomes a function of $\alpha$ alone and it is easy to determine its optimum. Taking into account that $0 \leq \alpha \leq 1$,  we find that the minimum failure probability is obtained when
\begin{widetext} 
\begin{equation}
\label{alpha1}
\alpha_{opt}= \left \{
\begin{array}{ll}
0  \;\; &  \mbox {if $ \;\;(n+2)\eta_1 \leq 2\eta_2/(n+1) $}  \\
\frac{n+1}{n}\left(1 - \sqrt{\frac{2\eta_2}{(n+1)(n+2)\eta_1}}\right)
         \;\; & \mbox{if $\;\;\frac{2\eta_2}{n+1} \leq (n+2)\eta_1 \leq 2(n+1)\eta_2$}\\       
 1  \;\; & \mbox{if $\;\;(n+2)\eta_1 \geq 2(n+1)\eta_2$} \ . \\       
\end{array}
\right.
\nonumber
\end{equation}
\end{widetext}
Substituting these values into (\ref{op41}) yields $\beta_{opt}$.    
Using $\alpha_{opt}$ and $\beta_{opt}$ in Eqs. (\ref{op11}) and (\ref{op31}) yields an explicit expression for the optimum detection operators.

If $(n+2)\eta_1 \leq 2\eta_2/(n+1)$, which implies that $\eta_1\leq 2/[2+(n+1)(n+2)]$, 
we have $\Pi_1^{\rm opt}=0$ and 
$\Pi_2^{\rm opt} = |\bar{v}_2\rangle \langle \bar{v}_2| + \sum_{i=3}^{n+2}|u_{i}\rangle \langle u_{i}|$, which means that the optimum measurement is a projection measurement on the subspace orthogonal to the span of $\rho_{1}^{\prime}$, i. e. a projection on its kernel.  On the other hand, for $(n+2)\eta_1 \geq 2(n+1)\eta_2$, i.e.\  $\eta_1 \geq 2(n+1)/(3n+4)$, the optimum measurement is 
a joint projection measurement on the kernels of $\rho_{1}^{\prime}$ and $\rho_{2}^{\prime}$, where
$\Pi_1^{\rm opt} = |\bar{u}_2\rangle \langle \bar{u}_2|$ and
$\Pi_2^{\rm opt} = \sum_{i=3}^{n+2}|u_{i}\rangle \langle u_{i}|$. In the intermediate parameter region the optimum measurement is a generalized measurement. The failure probability of these optimal  measurements can be summarized as
\begin{widetext} 
\begin{equation}
\label{Qfopt}
Q_{F}^{\prime} = \left \{
\begin{array}{ll}
\eta_{1} + \frac{\eta_2}{n+1} \;\; &  \mbox {if $ \;\; \eta_1 \leq \frac{2}{2+(n+1)(n+2)}$}  \\
\frac{\eta_{1}}{2} + \frac{\eta_{2}}{n+2} + \sqrt{\frac{2\eta_{1} \eta_{2}}{(n+1)(n+2)}}
         \;\; & \mbox{if $\;\;\frac{2}{2+(n+1)(n+2)} \leq \eta_1 \leq \frac{2(n+1)}{2(n+1)+(n+2)}$}\\       
 \eta_{1}\frac{n+2}{2n+2}+\eta_{2}\frac{2}{n+2}  \;\; & \mbox{if $\;\;\eta_1 \geq \frac{2(n+1)}{2(n+1)+(n+2)}$} \ . \\       
\end{array}
\right.
\nonumber
\end{equation}
\end{widetext} 
We notice the the above expressions reduce to the corresponding expressions of Sec. IV.A for $n=1$, as they should. Then, as in that section, it is also true here that the benefit of performing the generalized measurement is only marginal. To see the closeness of the best PVM (projective valued measurement) to the optimal POVM we compare their performance in several ways. The two PVMs (first and last line in (\ref{Qfopt})) deliver the same result for $\eta_{1}=2/(n+4)$. In fact, the reduction of the POVM failure probability (middle line) compared to those of the projective measurements is largest for this value of $\eta_{1}$. In Fig. 3 we display the PVM and POVM failure probabilities for this value of $\eta_{1}$ as a function of $n$. We see that the two curves remain close together for all values of $n$. 
\begin{figure}[ht]
\begin{center}
\epsfig{file=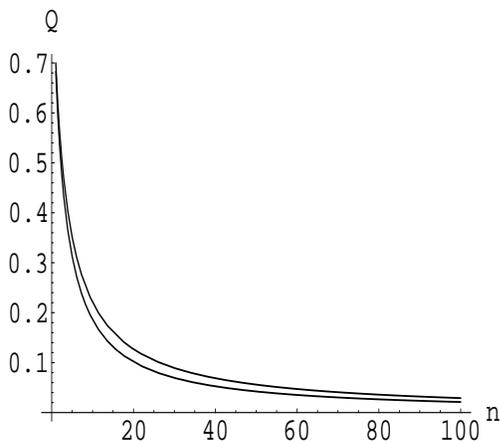,height=6cm,width=7cm}
\caption 
{Comparison of the optimum performances of the failure probabilities of the projective measurements (upper curve) and the failure probability of the POVM (lower curve) vs. the number of copies $n$ for the value of $\eta_{1}=2/(n+4)$, when their difference is the largest for the unambiguous discrimination of one known state from one unknown state when $n$ copies of the date state are provided.}
\end{center}
\end{figure}
The difference between these two curves as a function of $n$ reaches a maximum, however. It is maximal for $n=5$ as displayed in Fig. 4.
\begin{figure}[ht]
\begin{center}
\epsfig{file=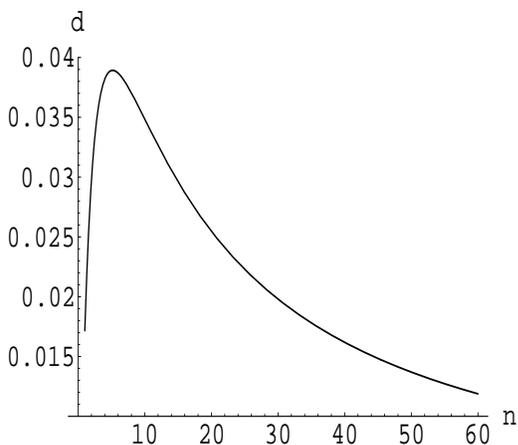,height=6cm,width=7cm}
\caption 
{The difference between the upper curve and the lower curve in Fig. 3 as a function of $n$. The difference between the performance of the PVM and POVM is maximum for $n=5$.}
\end{center}
\end{figure}
Finally, in Fig. 5 we display the ratio of the PVM failure probability to the POVM failure probability as a function of $n$. Asymptotically, the POVM outperforms the PVM by 50\%, their ratio tending to the limiting value of $1.5$. However, as we see from the figure, one needs about a $1000$ copies of the data state to reach the asymptotic region. Since the difference is maximal for five copies we can conclude that one does not need more than five copies in order to demonstrate performance enhancement due to the optimal POVM. 
\begin{figure}[ht]
\begin{center}
\epsfig{file=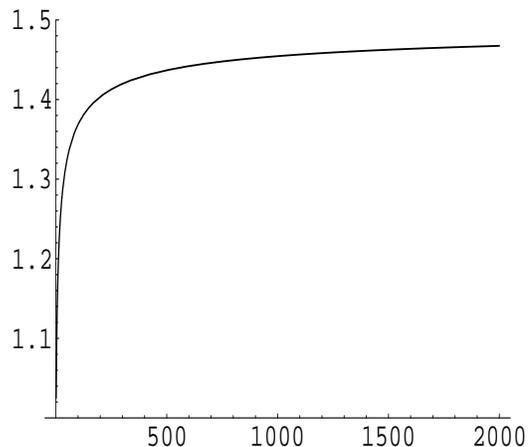,height=6cm,width=7cm}
\caption 
{The ratio of the upper and lower curve in Fig. 3 as a function of $n$. Asymptotically the failure rate of the PVM is 50\% higher than that of the POVM.}
\end{center}
\end{figure}

To close this section we also notice that, in agreement with the general relation derived in 
\cite{HB2}, the optimal POVM failure probability is always more than twice as large as the 
minimum error probability $P_E^{\prime}$ of the previous subsection.

\section{Conclusion}
We have described a number of quantum devices that discriminate
between two quantum states.  We do not possess complete information 
about the states to be discriminated.  Our devices
have two inputs, one for the qubit whose identity is to be determined,
and the other for the copies of one or both of the possible states 
that it can be in.  In the case that only one of the states is
provided, it is assumed that the other state is known, and this
knowledge is built into the device.  The states sent into the second
input can be regarded as a program.  To change the set of states 
between which we are discriminating, we do not have to change the 
device, but merely supply it with a different program.

We want to point out a striking feature of the programmable state
discriminators in which copies of both of the states to be 
discriminated are provided.   Neither the optimal detection
operators nor the boundaries for
their region of validity depend on the unknown states. 
Therefore, these devices are {\it universal},
they will perform optimally for any set of unknown states. Only the
probability of success for fixed but unknown states will depend on the
overlap of the states.  
However, both this expression and its  average over
all possible inputs is optimal.

The devices described here demonstrate the
role played by \emph{a priori} information.  All of them have a smaller
success probability than one designed for a case in which we know 
both of the input states, and the device for two unknown input states 
has a smaller success probability than one designed for the case 
when we know one of the input states.  There is a trade off between 
flexibility and success probability.  The more of the information 
about the states that is carried by a quantum program, the smaller 
the probability of successfully discriminating between the states, 
but the larger the set of states for which the device is useful.  
This flexibility suggests that programmable discriminators will be 
useful as parts of larger devices that produces
quantum states that need to be identified.

We conclude our paper by summarizing what we know about programmable 
discriminators with quantum programs in which the programs consist of 
copies of the
states to be discriminated.   The most general problem of this type is
when we have $n_A$ copies of the state of the program system $A$, $n_C$ 
copies of the state of the program system $C$, and $n_B$ copies of the 
state of the data system $B$. In this case, the task is to discriminate 
two input states
\begin{eqnarray}
|\Psi_{1}^{in}\rangle & = & |\psi_{1}\rangle_{A}^{\otimes 
n_A}|\psi_{1}\rangle_{B}^{\otimes n_B}
|\psi_{2}\rangle_{C}^{\otimes n_C} \ , \nonumber \\
|\Psi_{2}^{in}\rangle & = & |\psi_{1}\rangle_{A}^{\otimes 
n_A}|\psi_{2}\rangle_{B}^{\otimes n_B}
|\psi_{2}\rangle_{C} ^{\otimes n_C}\ ,
\end{eqnarray}
where the subscripts $A$ and $C$ refer to the program registers ($A$ 
contains
$|\psi_{1}\rangle$ and $C$ contains $|\psi_{2}\rangle$), and the 
subscript $B$
refers to the data register.  Our goal would be to optimally 
distinguish between
these inputs,
keeping in mind that one has no knowledge of $|\psi_{1}\rangle$ and 
$|\psi_{2}\rangle$
beyond their \emph{a priori} probabilities.  The problem in which the 
numbers of copies of the program states are equal and greater than one, 
but we have only
one copy of the data state is solved for equal \emph{a priori} probabilities \cite{hayashi2}.  The problem in which we have only one copy of each program state, but an arbitrary 
number of copies of the data state has been solved here.  The general problem remains open.

\begin{acknowledgments}
This research was partially supported by the European Union projects
CONQUEST and QAP, by a PSC-CUNY Grant and by the Slovak Academy
of Sciences via the project CE-PI (I/2/2005) and the project APVT-99-012304.  JB and VB are grateful for the hospitality extended to
them during their stay with the Department of Quantum Physics of Prof. Wolfgang Schleich
at the University of Ulm and to the Alexander von Humboldt Foundation for financial support.
JB also acknowledges the hospitality extended to him during his visit at
the Nanooptics group of Prof. Oliver Benson at the Humboldt University Berlin.
 
\end{acknowledgments}

\bibliographystyle{unsrt}

\end{document}